%% file: specfs.tex
\documentclass[letterpaper,twocolumn,10pt]{article}
\usepackage[available,functional,reproduced]{usenixbadges}
\usepackage{usenix-2020-09}

\usepackage{tikz}
\usepackage{amsmath}
\usepackage{tikz}
\usepackage{amsmath}
\usepackage[normalem]{ulem}
\usepackage{endnotes,microtype,xspace,fancyvrb,multirow}
\usepackage{xcolor}
\usepackage{xspace}
\usepackage{balance}
\usepackage{pifont}
\usepackage{anyfontsize}
\usepackage{booktabs}
\usepackage{fp}
\usepackage{caption}
\usepackage[frozencache, cachedir=.]{minted}
\usepackage{listings}
\usepackage{balance}
\usepackage{textcomp}
\usepackage{stmaryrd}
\usepackage{enumitem}
\usepackage{array}
\usepackage{wrapfig}
\usepackage{makecell}
\usepackage{stfloats}

\usepackage{subcaption}
\usepackage{titlesec}
\usepackage{tabularx}
\usepackage[T1]{fontenc}
\usepackage{bbding}  
\usepackage{url}
\usepackage{colortbl}

\usepackage{layout}
\usepackage{multicol}
\usepackage{csquotes}
\usepackage{tabularx}

\usepackage{booktabs}
\usepackage{multirow}

\usepackage{adjustbox}

\usepackage{xparse}
\usepackage{mathtools} 

\usepackage{amssymb}

\newcommand{\sys}{\textsc{SpecFS}\xspace}
\newcommand{\spec}{\textsc{SysSpec}\xspace}
\newcommand{\compiler}{SpecCompiler\xspace}
\newcommand{\validator}{SpecValidator\xspace}
\newcommand{\ass}{SpecAssistant\xspace}

\newcommand{\REVISE}[1]{#1}

\newcommand{\TODO}[1]{\textcolor{red}{TODO: #1}}

\definecolor{hjkcolor}{RGB}{85,107,47}

\newcommand{\codeword}[1]{\textcolor{black}{$\mathsf{#1}$}}
\newcommand{\myparagraph}[1]{\noindent\textbf{#1}}

\DeclareCaptionFont{myfont}{\fontsize{9.4pt}{11.28pt}\selectfont}
\DeclareCaptionFont{subfigfont}{\fontsize{9pt}{10.8pt}\selectfont}
\titlespacing*{\subsection}{2pt}{*1}{*1}
\titlespacing*{\subsubsection}{2pt}{*1}{*1}

\definecolor{diaozi}{RGB}{93,49,49}
\definecolor{hightcode}{rgb}{1.0,0.13,0.32}
\definecolor{hightcomment}{RGB}{205, 92, 92}
\definecolor{myRubineRed}{RGB}{209, 0, 86}

\renewcommand{\tiny}{\fontsize{6pt}{7pt}\selectfont}

\definecolor{diaozi}{RGB}{93,49,49}
\definecolor{hightcode}{rgb}{1.0,0.13,0.32}
\definecolor{myRubineRed}{RGB}{209, 0, 86}
\microtypecontext{spacing=nonfrench}

\lstdefinelanguage{CoqStyle}{
    morekeywords={Require, Import, Section, Variable, Hypothesis, 
                  Theorem, Proof, Qed, Definition, Inductive,
                  Fixpoint, match, end, with},
    sensitive=true,
    morecomment=[s]{(*}{*)},
    morestring=[d]{"}
}

\begin{document}

\date{}


\title{Sharpen the Spec, Cut the Code: A Case for Generative File System with \spec}

\author{\rm Qingyuan Liu,\; Mo Zou$$,\; Hengbin Zhang,\; Dong Du,\; Yubin Xia,\; Haibo Chen\\
{\normalsize {Institute of Parallel and Distributed Systems}} \\
{\normalsize {Shanghai Jiao Tong University}} 
}

\maketitle




\input{abstract}

\input{intro}

\input{motiv}

\input{overview}
\input{design}
\input{design-fs}
\input{eval}

\input{future}
\input{relat}

\input{concl}
\input{ack}








\bibliographystyle{plain}

\bibliography{ref}
\input{appendix}





\end{document}

%% file: abstract.tex
\begin{abstract}


File systems are critical OS components that require constant evolution to support new hardware and emerging application needs.
However, the traditional paradigm of developing features, fixing bugs, and maintaining the system incurs significant overhead, especially as systems grow in complexity.
This paper proposes a new paradigm, \emph{generative file systems}, which leverages Large Language Models (LLMs) to generate and evolve a file system from prompts,
effectively addressing the need for robust evolution.
Despite the widespread success of LLMs in code generation, attempts to create a functional file system have thus far been unsuccessful,
mainly due to the ambiguity of natural language prompts.

This paper introduces \spec, a framework for developing generative file systems.
Its key insight is to \emph{replace ambiguous natural language with principles adapted from formal methods}. 
Instead of imprecise prompts, \spec{} employs a multi-part \emph{specification} that accurately describes a file system's functionality, modularity, and concurrency. 
The specification acts as an unambiguous blueprint, guiding LLMs to generate expected code flexibly.
To manage evolution, we develop a \emph{DAG-structured patch} that operates on the specification itself, enabling new features to be added without violating existing invariants. Moreover, the \spec toolchain features a set of LLM-based agents with mechanisms to mitigate hallucination during construction and evolution.
We demonstrate our approach by generating \sys, a concurrent file system.
\sys demonstrates equivalent level of correctness to that of a manually-coded baseline across hundreds of regression tests.
We further confirm its evolvability by seamlessly integrating 10 real-world features from Ext4.
Our work shows that a specification-guided approach makes generating and evolving complex systems not only feasible but also highly effective.

\end{abstract}

%% file: intro.tex
\section{Introduction}
\label{s:intro}

File systems are a cornerstone of modern operating systems, providing the critical abstractions for managing persistent data.
Their design is dictated by a symbiotic relationship with two forces: the characteristics of underlying storage hardware and the demands of ever-changing applications.
This relationship necessitates continuous evolution, driving the creation of specialized file systems like F2FS for flash memory~\cite{188454} and EROFS for read-only mobile scenarios~\cite{234892}.
Consequently, file system developers are in a perpetual cycle of adding features, optimizing performance, and resolving bugs to keep pace with innovation.

However, this evolution comes at a steep price.
To quantify this challenge, we conduct a longitudinal study of the Ext4 file system's development, analyzing all 3,157 commits from its inception in Linux 2.6.19 to the recent 6.15 release. 
Our analysis reveals that 82.4\% of all commits are dedicated to bug fixes and maintenance, which are a direct consequence of introducing new functionality (which accounts for only 5.1\% of total commits).
E.g., the recently merged ``fast commits'' feature~\cite{298507} required only 9 commits for its initial implementation,
while it triggered about 80 subsequent commits to address newly introduced bugs and maintain the code (\autoref{ssec:case-study}).
This demonstrates a common development cycle where the effort to stabilize new features far outweighs the initial implementation effort, placing a non-trivial burden on developers.

This motivates us to explore a new paradigm for file system design and development, \emph{generative file systems}, which leverages the capabilities of Large Language Models (LLMs)~\cite{chatgpt, liu2024deepseek, liu2024deepseek2, team2024gemini, bai2023qwen} to generate a complete file system and evolve it effectively.

Nevertheless,
although recent advancements in LLMs have demonstrated their profound capabilities in automated code generation~\cite{dong2025qimeng,cursor, copilot, qi2025intention,zhang2024codeagentenhancingcodegeneration, huang2024agentcodermultiagentbasedcodegeneration,wang2025coderagbenchretrievalaugmentcode,zhang2024autocoderover},
generating a complete and useful file system from natural-language prompts is profoundly challenging, if not impossible.
The intricate semantics of file systems, from ensuring concurrency correctness to carefully managing complex file structures,
are difficult to express unambiguously in prompts.
Consequently, to our knowledge, no prior LLM-based approach has successfully generated a complete, functional, and feature-rich file system.

Compared with descriptions with natural language, \emph{specifications}~\cite{stoica2024specificationsmissinglinkmaking, ntzik_et_al:LIPIcs.ECOOP.2018.4} usually provide a precise and machine-understandable language to
\emph{explicitly encode} these invariants (e.g., no orphan inodes after a crash),
creating an opportunity to guide LLMs toward generating expected code.
However, bridging the gap between specifications and LLM-based code generation is non-trivial.
We identify three technical challenges that our work, \spec, is designed to overcome:

\myparagraph{Challenge I: Specification semantic gap.}
A specification must be expressive enough to capture the multifaceted semantics of a file system, which natural language prompts fail to do.
This includes not only functional correctness but also non-functional properties critical for performance, such as on-disk layout choices (e.g., bitmap vs. linear scan for block allocation).
Furthermore, precisely specifying complex concurrency control is notoriously difficult. Attempting to describe both functional logic and fine-grained locking in a single, monolithic prompt can overwhelm an LLM, causing it to overlook subtle but critical details and produce code with concurrency bugs.

\myparagraph{Challenge II: Complex component composition.}
The finite context window of LLMs precludes generating an entire file system monolithically.
This necessitates a modular approach, which introduces significant composition challenges.
First, generating one module at a time, an LLM lacks the global context to ensure interface compatibility with other, yet-to-be-generated or pre-existing components, leading to integration errors.
Second, each individual change (with LLMs) introduced during the evolution process can potentially trigger cascading effects on other file system modules,
particularly those with existing dependencies.
For example, a seemingly local feature addition, like introducing extent~\cite{linux-ext4}, can trigger non-local changes by altering core data structures like the inode,
affecting any module that interacts with it and making manual dependency management intractable.

\myparagraph{Challenge III: Unreliable LLM capability.}
LLM-based code generation is inherently non-deterministic due to the hallucination~\cite{10.1145/3703155, xu2025hallucinationinevitableinnatelimitation} --- even identical specifications can yield different and potentially incorrect code outputs across generation attempts.
A naive ``generate-and-pray'' approach is unacceptable for system software.
Therefore, a robust framework cannot blindly trust the LLM especially for file systems;
it must incorporate a rigorous validation mechanism to ensure that generated code strictly adheres to the guiding specification, guaranteeing correctness despite the unreliability.

\begin{figure*}[t]
  \centering
 \setlength{\belowcaptionskip}{-10pt}
 \setlength{\abovecaptionskip}{0pt}
  \begin{minipage}[t]{0.9\linewidth}
      \centering
      \includegraphics[width=\textwidth]{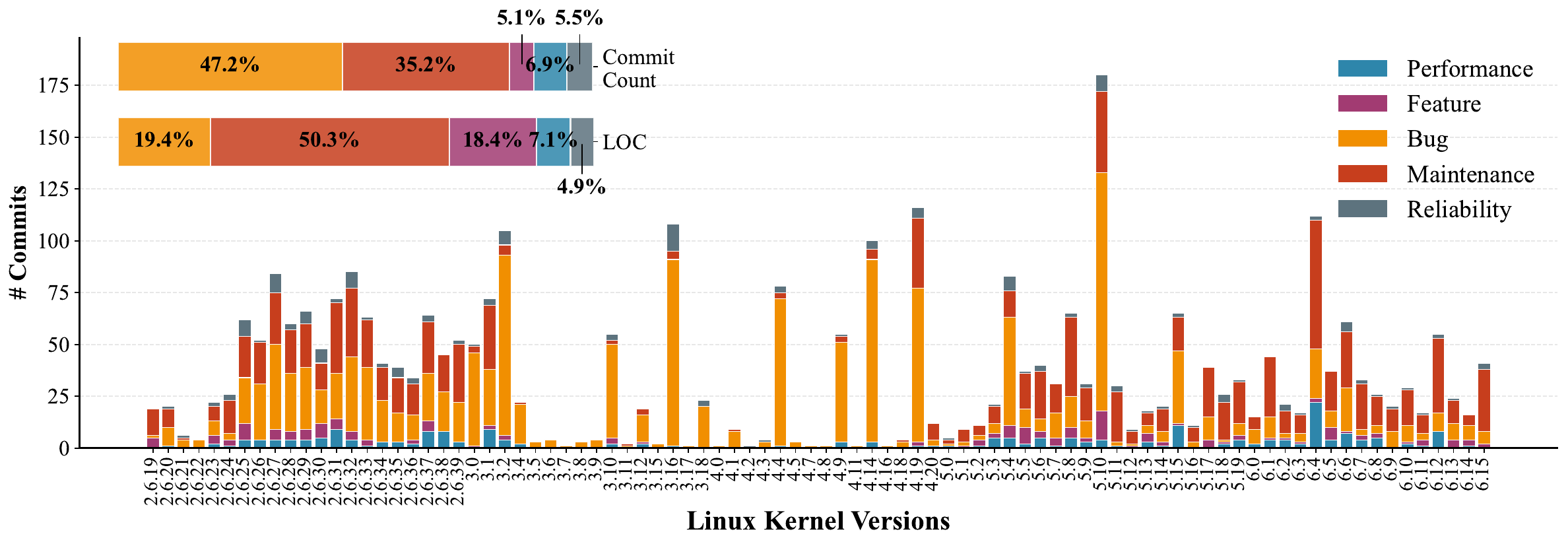}
      \footnotesize
    \end{minipage}
  \caption{\textbf{File system evolution with different types of patches}.
  }
  \label{fig:motiv-patch-type-evolution}
\end{figure*}

To address these challenges,
    our key insight is to \emph{replace ambiguous natural language with specifications following the principles adapted from successful practices of formal methods}~\cite{zou2019atomfs, chen2015using, 199293}.
Specifically, we introduce \spec{}, a framework for correct-by-construction file system generation built on three core techniques.

First, we design a \textbf{formal method-inspired specification} to precisely define a file system's behavior, moving beyond the ambiguity of natural language.
It holistically captures the FS design, spanning \textbf{Functionality}, defined through Hoare-logic based pre/post-conditions~\cite{zou2019atomfs, chen2015using} and invariants;
\textbf{Modularity}, which enforces a clean decomposition into modules with rely-guarantee~\cite{zou2019atomfs, Liang12popl} interfaces for correct composition;
and \textbf{Concurrency}, which makes locking protocols and ordering explicit to mitigate subtle bugs.

Second, \spec{} provides an \emph{LLM-based toolchain} that translates the high-level specification into an executable C implementation.
The toolchain contains three LLM-based agents: the \compiler{}, which systematically synthesizes C code from the specification using techniques like \emph{two-phase generation} (logic first, then concurrency);
the \validator{}, which rigorously validates the generated code against the specification and drives a \emph{retry-with-feedback} loop to autonomously correct LLM errors;
and then \ass which eases the development of specification.

Third, we introduce a mechanism for \textbf{principled evolution via spec patches}.
Instead of manually modifying C implementation,
developers add features by authoring a special patch, called \emph{spec patch}, to the high-level specification.
The \spec{} toolchain then automatically propagates this change, regenerating the implementation to ensure the new feature is correctly and consistently integrated.

Overall, the new paradigm with \spec shifts the developer's burden from low-level implementation to high-level design.
While this demands greater upfront design effort, akin to the safety discipline of Rust~\cite{safe-rust, bugden2022rustprogramminglanguagesafety},
the return is a file system that is far easier to maintain and evolve.


We utilize \spec to implement \sys{}, a complete FUSE-based (generative) file system specified entirely in specification. 
Our toolchain automatically generates a functional C-language implementation without any manual intervention.
We validate its correctness and expressiveness by successfully generating implementations of a previously verified file system, AtomFS~\cite{zou2019atomfs}.
Besides, we showcase its capability for complex feature integration by seamlessly evolving \sys with spec patches to support 10 novel features from Ext4,
including delayed allocation and file encryption.
We also highlight its benefits on performance: by applying Ext4-style delayed allocation method with a concise spec patch, 
\spec{} automatically regenerated the relevant modules, resulting in a 99.9\% data write reduction for xv6 compilation.

To our knowledge, \spec is the first framework that enables both the generation and principled evolution of end-to-end file systems like \sys,
shifting the developer's focus from writing brittle low-level code to crafting robust, high-level designs.
\spec and \sys is available at: \url{https://llmnativeos.github.io/specfs/}.

%% file: motiv.tex
\section{Characterizing the Evolution of File Systems}

File systems 
usually require a continuous evolution to adapt to new hardware features and emerging use cases.
This constant adaptation includes adding new features, resolving bugs, and optimizing for critical scenarios.
Understanding the intricate process of file system evolution
offers valuable insights for future design.
Prior work by L. Lu et al.~\cite{180726} in 2013 provided the first analysis of the Linux file system's evolution.
We continue the analysis with an extended study of evolution over a 20-year period (from 2005 to 2025),
and identify new implications that highlight the need to re-evaluate current file system paradigms,
particularly in the context of the new opportunities provided by large language models.

\subsection{The Anatomy of File System Change}
\label{ssec:anatomy-fs-change}

\myparagraph{Methodology.}
We choose Ext4 as our target because it is a mature and (still) widely-deployed file system with 20 years of evolution in real-world environments.
Ext4 was introduced in Linux 2.6.19 and has been continuously developed, incorporating thousands of patches related to performance, new features, and maintenance.
Our analysis is based on all 3,157 Ext4-related commits \REVISE{(i.e., individual patches rather than patch sets)} merged into the mainline Linux from version 2.6.19 to 6.15.
To understand the evolution, we categorize each commit using a classification scheme for FS patches adapted from prior work~\cite{evolutionLu2014}:
(1) Bug: fixing an existing bug,
(2) Performance: improving efficiency through new designs or optimizations,
(3) Reliability: enhancing the file system's robustness,
(4) Feature: implementing new functionalities,
and (5) Maintenance: refactoring code or improving documentation without changing semantic behavior.


\myparagraph{Implication-1: File systems consistently evolve.}
Our analysis of Ext4 patches across Linux versions, as shown in \autoref{fig:motiv-patch-type-evolution}, reveals a clear evolutionary trend.
Initially, during the early stages (Linux 2.6.19--3.4), the high number of changes reflects extensive work on new features, bug fixes, and maintenance.
The number of changes then decreases significantly from Linux 3.4 to 4.18 as the codebase matures.
However, a surprising trend emerges: changes increase steadily after Linux 4.19, peaking at Linux 5.10 --- more than a decade after Ext4's introduction.
We also observe occasional peaks in the number of changes even during the stable period (Linux 2.6.19--3.4), such as over 50 changes in Linux 3.10 and over 100 in Linux 3.16.
This result suggests that a long-lived file system like Ext4 is in \emph{a state of constant evolution}.

\begin{figure}[t]
 \setlength{\belowcaptionskip}{-10pt}
 \setlength{\abovecaptionskip}{0pt}
  \centering
  \includegraphics[width=0.48\textwidth]{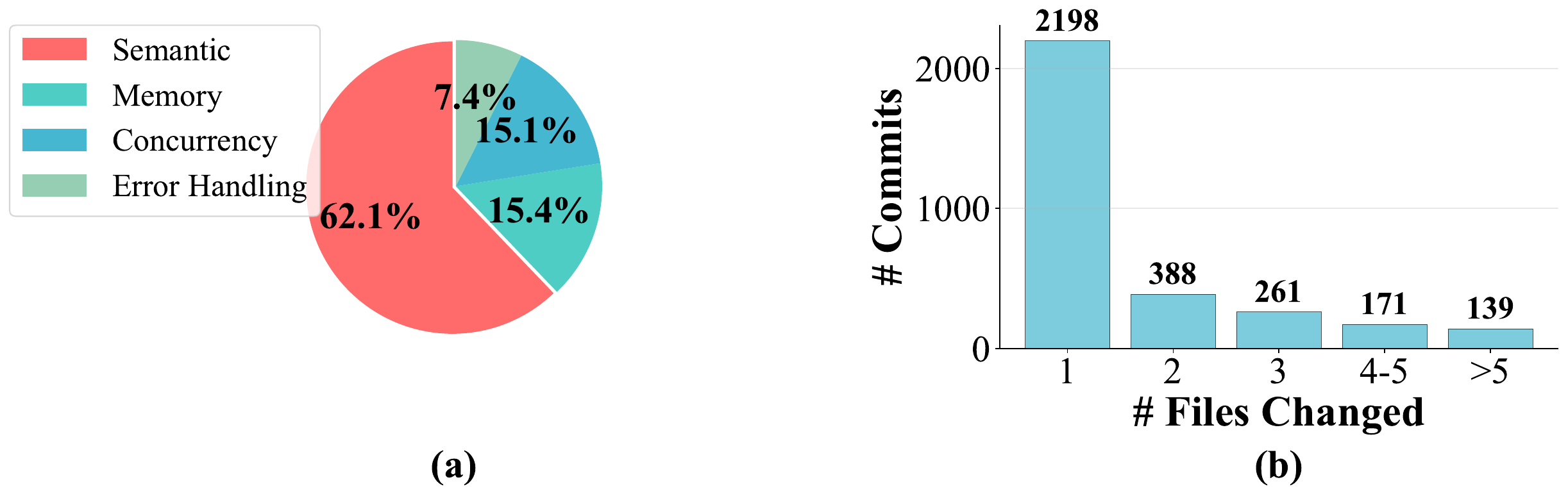}
  \caption{\textbf{(a) Distribution of bug type. (b) Distribution of changed files per commit. }
  }
  \label{fig:motiv-distribution}
\end{figure}

\myparagraph{Implication-2: Bug fixes and maintenance dominate a file system's lifetime.}
While users are often drawn to a file system by its key features, e.g., Ext4's use of extents,
our analysis reveals that non-feature patches dominate the file system's lifetime.
Specifically, 82.4\% of all commits focus on bug fixes and maintenance.
\autoref{fig:motiv-distribution}-a further shows the distribution of bug type, indicating that most bugs are semantic bugs. 
This finding presents an insight: although new features initially attract users, the long-term success and widespread adoption of a FS depend heavily on its continuous maintenance.
However, this high volume of maintenance work represents a significant and ongoing burden,
and current paradigms lack efficient solutions to handle it.


\myparagraph{Implication-3: Feature-related changes are non-trivial.}
While feature-related commits constitute only 5.1\% of the total changes, their impact is significant.
Our analysis of the code base, as shown in \autoref{fig:motiv-patch-type-evolution}, reveals that these commits account for 18.4\% of the total lines of code (LOC) changed.
Furthermore, we observe that the introduction of a new feature often acts as a catalyst for subsequent bug fixes and maintenance work.
A new feature introduces a new code base, which in turn necessitates follow-up commits to fix newly introduced bugs and maintain the code's integrity.
Consequently, despite their low commit count, feature-related changes are central to the overall evolution of a file system.

\begin{figure}[t]
 \setlength{\belowcaptionskip}{-10pt}
 \setlength{\abovecaptionskip}{0pt}
  \centering
  \includegraphics[width=0.45\textwidth]{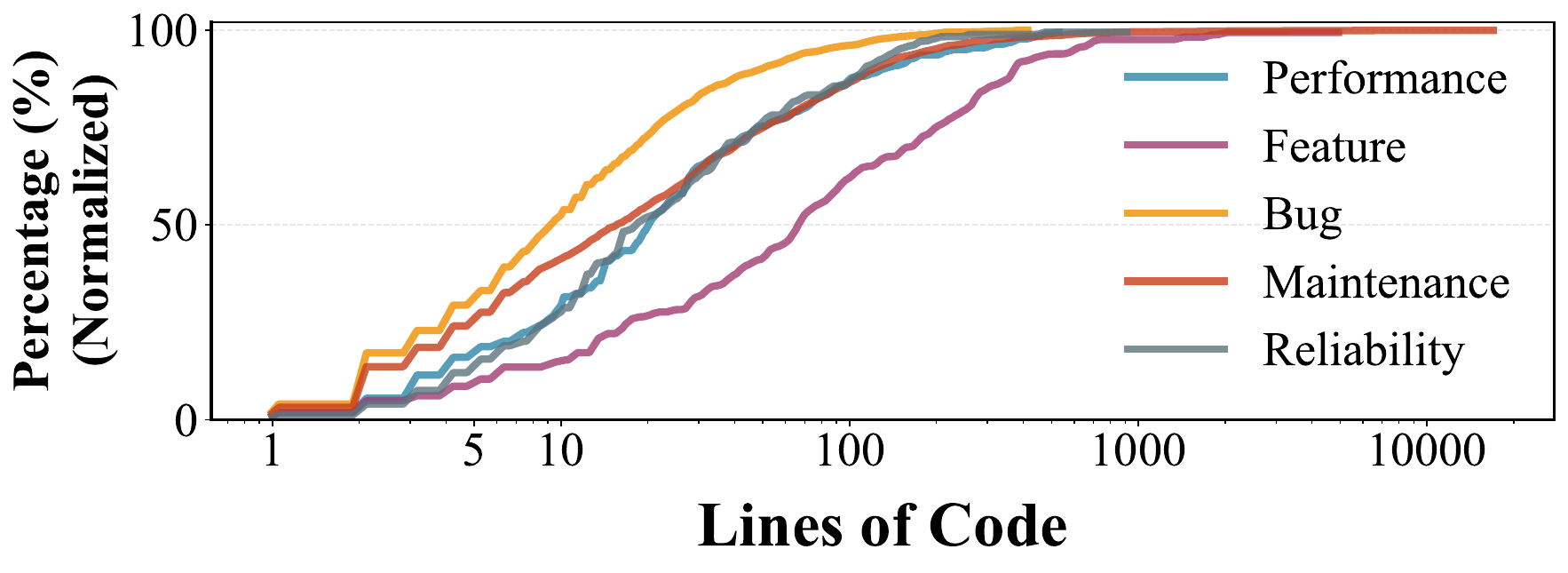}
  \caption{\textbf{Patch LOC size CDF}.
  }
  \label{fig:motiv-patch-size-cdf}
\end{figure}

\myparagraph{Implication-4: Evolution is taken in small steps.}
We measured the LOC for each patch and present the cumulative distribution function (CDF) in \autoref{fig:motiv-patch-size-cdf}.
Our analysis shows that evolution proceeds through small, frequent changes.
Specifically, about 80\% of all bug fixes involve fewer than 20 LOC.
While feature-related patches are generally larger, about 60\% of them still require fewer than 100 LOC.
Moreover, the vast majority of commits also modify only a single file, as shown in \autoref{fig:motiv-distribution}-b. 
These findings suggest that file system evolution is composed of these manageable, small-scale changes.

\subsection{Case Study: Evolution of Fast Commit}
\label{ssec:case-study}

We present a case study to illustrate how different types of patches are intertwined during the evolution. 
We use \emph{fast commit}~\cite{298507}, a hybrid journaling feature designed to optimize \texttt{fsync()}-intensive workloads, as an example.
Fast commit reduces I/O overhead and latency by using lightweight, logical commits, while periodically issuing full commits to maintain consistency.
It is one of the largest features introduced in Linux 5.10, comprising over 4,000 SLOC.

\myparagraph{Phase-1: Feature development.}
Our analysis covers 98 fast-commit-related patches from Linux kernel 5.10 to 6.15.
The initial implementation is concentrated in version 5.10, which includes 9 of the 10 feature-related commits.
These initial patches collectively introduce \texttt{jbd2} APIs for fast-commit support, initialization and recovery paths, and the main commit logic.
In total, these commits add over 4,000 lines of code spanning multiple core modules (e.g., inode, file, journal).
Despite affecting several modules, the modifications are carefully localized to minimize interference.
This modular design preserves the existing journaling mechanism and on-disk format while integrating the new fast-commit logic. 

\myparagraph{Phase-2: Bug fixes and stabilization.}
Development efforts shift to stabilization after the initial release.
Of the 55 bug-fix commits we have identified, over 65\% address semantic errors, e.g., misordered updates and incorrect handling of corner cases.
This high proportion of semantic bugs can be attributed to the feature's complexity and its deep integration with other Ext4 components.
We classify these bugs as either \textbf{internal} (within the fast-commit logic) or \textbf{cross-module} (from interactions with other parts of Ext4).
\autoref{fig:code:fast-commit} shows two examples.
The first illustrates an internal bug where an early return path omitted necessary cleanup operations, leading to lost metadata updates.
The second shows a cross-module bug where newly defined flags conflicted with existing journal checksum bits, requiring a redefinition of mount macros to resolve the collision.
These examples highlight the inherent difficulty of integrating features that span multiple modules,
underscoring the need for careful reasoning about global system invariants.

\input{code-fast-commit}

\myparagraph{Phase-3: Code maintenance.}
Alongside bug fixes, 24 maintenance commits (totaling 1,080 lines) are applied to refactor and document the fast-commit implementation. 
Examples include: (1)~\textbf{Refactoring for readability}, where logic for updating statistics is extracted into a dedicated function, \texttt{ext4\_fc\_update\_stats()},
to simplify the main \texttt{ext4\_fc\_commit()} pathway.
(2)~\textbf{API clarification}, which involves enhancing flag descriptions in both the source code and documentation to prevent misconfiguration.

The lifecycle of the fast-commit feature exemplifies a core challenge in the evolution of file systems:
the initial integration of a feature is usually followed by a long tail of numerous,
fine-grained fixes and refactoring efforts that are crucial for stability and long-term maintainability.



\subsection{A New Paradigm: Generative File System}
\label{ssec:opportunity}

The previous analysis reveals that FS development involves not only the inherent complexity of feature development but also the long tail of maintenance and bug fixes. 
This entire process is laborious and error-prone when performed manually in low-level C code.
Luckily, the recent advances in Large Language Models (LLMs) offer an opportunity to address this challenge.
LLMs' proficiency in code generation, refactoring, and reasoning is well-suited for the high volume of maintenance and bug-fix tasks that consume the majority of developer effort.
Furthermore, the scope of code when evolving a file system could typically fit within the context windows of modern LLMs (Implication IV, \textsection\ref{ssec:anatomy-fs-change}), making an automated approach technically feasible.

To this end, we propose the new paradigm of \emph{generative file system}, i.e., leveraging LLMs to generate and evolve a file system.
The central thesis of our work is that:
\emph{the path forward is not to replace developers with LLMs, but to elevate their role by changing how they express design}.

\input{tab-relat}

\myparagraph{Limitations of prior works.}
Although abundant prior works achieve automatic code generation, their inherent limitations prevent us from leveraging them to achieve the paradigm for complex file systems, as generalized in \autoref{tab:relat}. 
These works can be broadly categorized into two types. 
First, some works focus on generating complete code logic from scratch (``from $0$ to $N$''). 
However, this category of work struggles to cope with the complexity of file systems, whose modules exhibit intricate interactions and dependencies. 
Consequently, such approaches either produce only simple code (e.g., frontend pages) or are limited to specific single-module tasks~\cite{sun2024cloverclosedloopverifiablecode,dong2025qimeng}, 
such as generating tensor program implementations for different hardware~\cite{dong2025qimeng}, 
generating unit tests~\cite{sotiropoulos2024thalia, chen2022codetcodegenerationgenerated}, mathematical library functions~\cite{briggs2024megalibm}, or performing bit-vector synthesis~\cite{ding2024dryadsynth}. 

Second, other approaches usually focus on evolution, i.e., modifying existing code (``from $N$ to $N+1$'').
They may utilize methods like Retrieval-Augmented Generation (RAG)~\cite{wang2025coderagbenchretrievalaugmentcode,qi2025intention} or Agents~\cite{cursor, copilot, qi2025intention, zhang2024autocoderover,zhang2024codeagentenhancingcodegeneration, huang2024agentcodermultiagentbasedcodegeneration} to enhance general code generation capabilities.
However, these methods largely rely on natural language descriptions (e.g., document strings, GitHub issues~\cite{zhang2024autocoderover} or code review comments~\cite{qi2025intention}) for the desired program logic.
Even though some methods attempt to enhance the LLM's capability to understand intents expressed in natural language~\cite{qi2025intention}, ambiguity remains unavoidable: 
one cannot simply instruct an LLM to ``avoid race conditions'' and expect a correct outcome. 
Furthermore, these methods often include all project-related code in the context, requiring the Agent to autonomously decide how to retrieve information from this context. 
Such a burden is uncontrollable, and an excessively long context can potentially degrade the quality of code generation.

\myparagraph{Insight and challenges.}
Our key insight is that we could guide the LLM using principles derived from formal methods, rather than imprecise natural language prompting.
Such a formally structured specification is expected to address the challenges outlined in \textsection\ref{s:intro} simultaneously. 
First, a semantic gap arises when specifying module logic, necessitating resolution of (i) semantic ambiguities, (ii) deep domain knowledge awareness, and (iii) thread-safe requirements. 
Second, complex component composition demands careful consideration of inter-module dependencies and mitigation of cascading effects induced by each evolutionary patch across other modules. 
Third, the inherent unreliability of LLM capabilities poses a critical barrier to robust system generation.

%% file: code-fast-commit.tex
\begin{figure}
	\setlength{\belowcaptionskip}{-10pt}
	\setlength{\abovecaptionskip}{0pt}
	\centering
	\setlength{\columnsep}{0.2cm}
	\begin{minipage}[t]{0.950\linewidth}
		\begin{minted}[
			numbersep=\parindent,
			escapeinside=||,
			breaklines,
			fontsize=\fontsize{8pt}{9pt}\selectfont,
			tabsize=0
		]{C}
//  1. Internal logical error.
|\textcolor{red}{\textbf{--}}| if (unlikely(error))
|\textcolor{green}{\textbf{++}}| if (unlikely(error)) {
|\textcolor{green}{\textbf{++}}|    ext4_fc_stop_update(inode);
      return error;
|\textcolor{green}{\textbf{++}}| }
//  2. Cross-module collision.
|\textcolor{red}{\textbf{--}}| #define EXT4_MOUNT2_DAX_INODE 0x000010
   #define EXT4_MOUNT2_JOURNAL_FAST_COMMIT 0x000010
|\textcolor{green}{\textbf{++}}| #define EXT4_MOUNT2_DAX_INODE 0x000040
		\end{minted}
	\end{minipage}
	\caption{\textbf{Two Example Patches in \texttt{fast-commit}.}}
	\label{fig:code:fast-commit}
\end{figure}

%% file: tab-relat.tex

\begin{table}[t]
    \centering
    \caption{\textbf{Prior code generation methods.}\textit{``$0$ to $N$'' denotes generating code from scratch, while ``$N$ to $N+1$'' refers to generating code based on existing code, representing two categories of current work.}}
    \newcolumntype{C}[1]{>{\centering\arraybackslash}p{#1}}
    \newcolumntype{L}[1]{>{\raggedright\arraybackslash}p{#1}}
    \label{tab:relat}
    \footnotesize

    \setlength{\tabcolsep}{1pt}
    \resizebox{0.48\textwidth}{!}{%
    \begin{tabular}{l|l|ccc|l}
        \hline
        
        \textbf{Type} & \textbf{Prior works}                                                                  & \textbf{Precise}                                                 & \textbf{Modular}                                           & \textbf{Concurrent}   & \textbf{Specification}                                           \\ \hline
        \multicolumn{1}{l|}{\multirow{3}{*}{\begin{tabular}[c]{@{}l@{}}$0$ to\\$N$\end{tabular}}} & Copilot~\cite{copilot} & \textcolor{red}{\ding{55}} & \textcolor{green}{\ding{51}}  & \textcolor{red}{\ding{55}} & Natural Language \\
        \multicolumn{1}{l|}{} & Clover~\cite{sun2024cloverclosedloopverifiablecode}        & \textcolor{green}{\ding{51}}   & \textcolor{red}{\ding{55}} & \textcolor{red}{\ding{55}}          & Docstring + Annotation \\
        \multicolumn{1}{l|}{} & Qimeng~\cite{dong2025qimeng}                        & \textcolor{green}{\ding{51}}   & \textcolor{red}{\ding{55}} & \textcolor{red}{\ding{55}}          & Programming Language\\
        \hline
        \multicolumn{1}{l|}{\multirow{3}{*}{{\begin{tabular}[c]{@{}l@{}}$N$ to\\ $N+1$\end{tabular}}}} & AutoCodeRover~\cite{zhang2024autocoderover}                & \textcolor{red}{\ding{55}} & \textcolor{green}{\ding{51}}& \textcolor{red}{\ding{55}}   & Github Issue\\
        \multicolumn{1}{l|}{} & CodeAgent~\cite{zhang2024codeagentenhancingcodegeneration} & \textcolor{red}{\ding{55}} & \textcolor{green}{\ding{51}}  & \textcolor{red}{\ding{55}}   & Natural Language\\
        \multicolumn{1}{l|}{} & ``Intention''~\cite{qi2025intention} & Half & \textcolor{red}{\ding{55}}  & \textcolor{red}{\ding{55}}   & Natural Language\\

        \hline
        \multicolumn{1}{l}{}  & \textbf{\sys}                                                     & \textcolor{green}{\ding{51}} & \textcolor{green}{\ding{51}} & \textcolor{green}{\ding{51}} & \spec + Toolchain \\
        \hline
        \end{tabular}}
\end{table}

%% file: overview.tex
\begin{figure*}[t]
  \centering
  \includegraphics[width=0.95\textwidth]{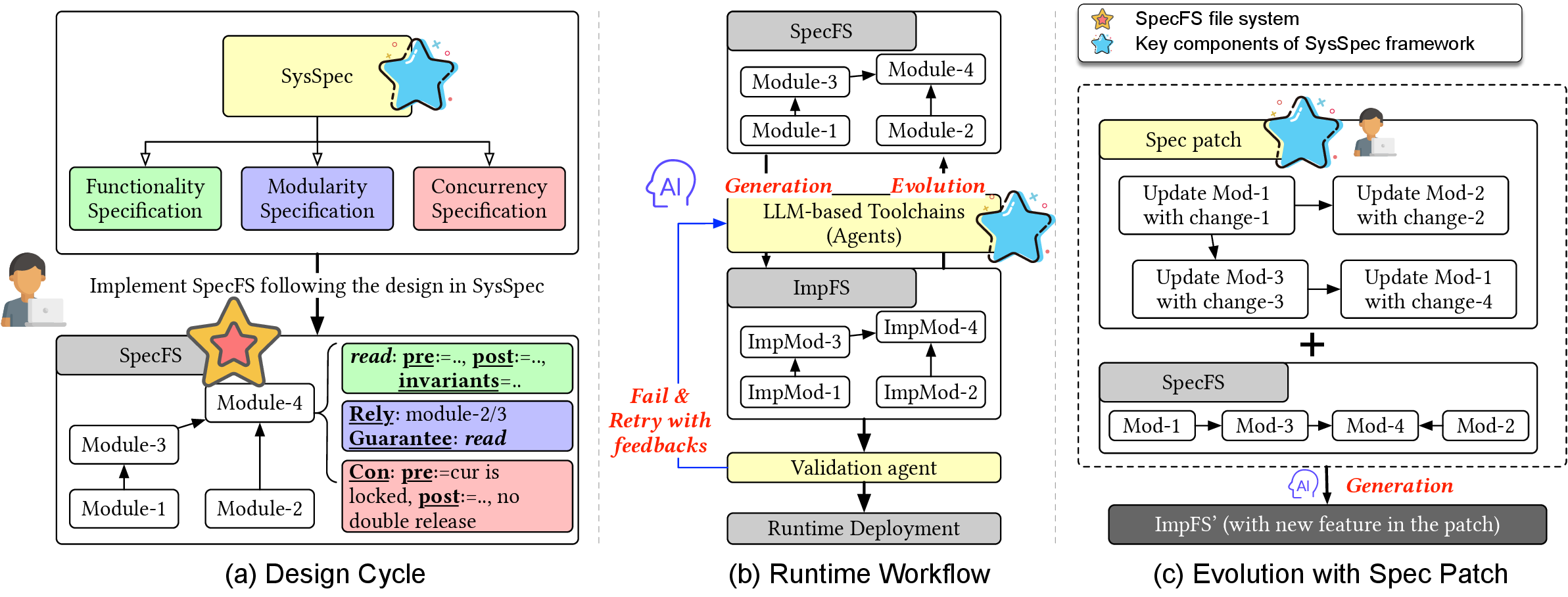}
  \setlength{\belowcaptionskip}{-15pt}
  \setlength{\abovecaptionskip}{0pt}
  \caption{\textbf{Design overview.} 
  (a) Developers write a file system (\sys) using a structured specification language that defines functionality, modularity, and concurrency. 
  (b) An LLM-based toolchain generates a low-level implementation (\texttt{ImpFS}) from the specification and uses a validation agent to ensure correctness.
  (c) The file system evolves by applying a high-level \textit{spec patch} to \sys, after which the toolchain regenerates the implementation to include the new features or fixes.}
  \label{fig:design-overview}
\end{figure*}

\section{Design Overview}

This paper presents \spec{},
an end-to-end framework for developing specification-based generative file systems.
\spec carefully applies principles from formal methods to create structured, high-level specifications.
While not strictly formal, these specifications provide a sufficiently precise blueprint to effectively guide LLMs in generating and evolving complex FS implementations.

\myparagraph{The \spec{} specification.}
At the core of our framework is a multi-part specification that captures a developer's design intent, as shown in \autoref{fig:design-overview}-a.
It consists of:
(1)~\textbf{Functionality specifications}, which use concepts like Hoare logic (pre/post-conditions) and invariants to describe the behavior of individual modules.
(2)~\textbf{Modularity specifications}, which decompose the system into distinct components and use a rely-guarantee discipline~\cite{10.1145/1480881.1480922} to ensure they can be composed correctly and developed independently.
(3)~\textbf{Concurrency specifications}, which explicitly define locking protocols and other concurrency-related behaviors that are notoriously difficult for LLMs to infer on their own.

\spec shifts the developer's role from writing low-level C code to authoring a high-level design.
This paradigm usually demands a greater upfront investment in design, much like the safety guarantees in Rust require more care than traditional C programming.
However, the payoff is significant: once specified, the file system is more robust, easier to evolve, and better suited for diverse scenarios
because the developer is forced to reason about the design concretely.

Besides, \spec{} streamlines evolution using \textbf{spec patches}, as shown in \autoref{fig:design-overview}-c. 
Instead of manually modifying thousands of lines of C code to add a feature, a developer writes a patch containing either new specifications or modifications to an old one.
Our toolchain then applies this patch and regenerates the low-level implementation, ensuring the change is propagated correctly throughout the system.

\myparagraph{The \spec toolchain.}
\spec{} provides an LLM-based toolchain that translates the high-level specification into an executable implementation. As illustrated in \autoref{fig:design-overview}-b, this toolchain includes agents:
The \textbf{\compiler{}} agent translates the specification (\sys) into a low-level C-language implementation (\texttt{ImpFS}). To manage the current limitations of LLMs, it uses techniques like two-phase prompting to handle complex logic.
The \textbf{\validator{}} agent ensures the correctness of the generated code, employing a retry-with-feedback loop to automatically identify and correct errors.
With the designs of \spec{}, our evaluation shows significant correctness improvements for complex operations compared to naive prompting (up to 34.4\%).
We also provide \textbf{\ass} to ease the development of specification.

\myparagraph{Case study: \sys.}
With \spec, we design and implement \sys{}, a complete file system written entirely in the \spec{} specification language. Our toolchain can automatically generate a functional C-language implementation of \sys{} without human intervention. Through a series of case studies, we demonstrate that \sys{} can seamlessly evolve via spec patches to support sophisticated features found in Ext4, e.g., extents or delayed allocation.

%% file: design.tex
\section{\spec Framework}
\label{s:design}

\subsection{Functionality Specification}
\label{subs:design-function}

The functionality specification defines the behavior of a module by describing its state transitions.
A module is a collection of related state variables and functions.
The specification is built upon three components, inspired by formal methods.

First, \textbf{pre- and post-conditions}, following Hoare Logic, define the contractual obligations for each function by specifying the required state before execution and the guaranteed state upon completion.
Second, \textbf{invariants} are properties that must hold true across all state transitions, ensuring the module's integrity.
\REVISE{Such Hoare-logic-based specification approach inherently avoids the need for explicitly constructing the entire state space, 
thereby directly circumventing the problem of state space explosion.}
Finally, a \textbf{system algorithm} outlines the high-level logic for how a function should achieve its state transition, guiding the LLM's implementation strategy.
Rather than always providing a complete system algorithm, 
our experience shows that a high-level \textbf{intent} is often sufficient. 
The intent can be regarded as a lightweight system algorithm that, 
expressed in natural language, guides the LLM in generating the desired implementation.

Not all of these components are required for every module.
We find that the necessary level of detail scales with complexity.
For straightforward modules (\textbf{Level 1}), pre/post-conditions 
and (sometimes) invariants are often sufficient.
As the logic becomes more intricate (\textbf{Level 2}), adding an intent description is recommended to clarify the design.
For the most complex cases involving highly optimized designs (\textbf{Level 3}),
providing an explicit algorithmic description becomes essential, as it is unreasonable to expect an LLM to derive such logic from scratch.



\input{code-ins-spec}




\myparagraph{Hoare logic for file system synthesis.}
To specify function behavior, we adapt the classic Hoare logic formalism of \{P\}C\{Q\} using pre- and post-conditions,
but we sidestep the high complexity of full formal verification.
Each function is annotated with \textbf{pre-conditions} that define required system states and \textbf{post-conditions} that guarantee specific state transitions and return values.
\autoref{fig:hoare-spec} shows \texttt{atomfs\_ins}, an internal function of AtomFS~\cite{zou2019atomfs} that implements \texttt{mknod}/\texttt{mkdir}. 
The pre-condition describes the validness of the parameters.
The post-condition describes the state after the operation. 

\REVISE{Our approach diverges from traditional formal methods in two key aspects.
First, an LLM enforces adherence to this logic during code generation, replacing the role of a formal theorem prover.
Second, to balance the comprehensibility and unambiguity of the specification, 
our specifications are expressed in structured natural language augmented with type annotations rather than pure mathematical logic, thereby making them more accessible.
\spec augments the semantic precision of natural language with a structured organization for the specifications (e.g., sections for functionality, modularity, and concurrency),
and uses mathematically disciplined natural-language expressions to ensure that the intended semantics remain precise and unambiguous while keeping the specification accessible. 
For example, the specification states that \textit{the file size equals $max(\text{old\_size}, \text{offset}+\text{len})$}, rather than that \textit{the write updates the file size if necessary}. 
These designs significantly limit the potential for misinterpretation.
}


\myparagraph{Invariant-guided specification.}
In addition to per-function contracts, developers specify system-wide \textbf{invariants} that describe properties valid across all states during execution. For example, an invariant may state that:
\begin{lstlisting}[language=C,basicstyle=\small\ttfamily]
[Invariant] any modification of an inode must 
occur while holding the corresponding lock
\end{lstlisting}
Such invariants define constraints that functions must respect throughout their execution, and cannot be fully expressed using only local pre- and post-conditions.
For another example,
the root existence invariant in \autoref{fig:hoare-spec} allows 
the generated code to safely omit null checks when accessing the root.

\myparagraph{System algorithm.}
While pre- and post-conditions define what a function must accomplish, they do not specify how.
Our experience shows this is insufficient for performance-critical systems, as an LLM might generate an implementation that satisfies the specification but is highly inefficient.
E.g., given a specification for a \codeword{sort()} function,
an LLM could correctly generate a bubble sort ($\mathcal{O}(n^2)$) just as easily as a quicksort ($\mathcal{O}(n \log n)$). To address this, the \textbf{system algorithm} component allows developers to explicitly outline the method for achieving a state transition.
This provides crucial guidance, ensuring the LLM implements a performant algorithm, such as using lock coupling for fine-grained concurrency instead of a coarse-grained global lock.

For example, in the functionality specification of \texttt{atomfs\_rename}, 
we define the algorithm in three phases: (1) traversing the common path, 
(2) traversing the remaining path and (3) checks and operations.
We explicitly specify the fine-grained locking scheme used in these phrases, 
which enables the correct generation of \texttt{atomfs\_rename}, a function that is both highly complex and prone to deadlock.

\myparagraph{Intent.}
LLMs are trained on vast codebases and can often produce highly optimized code when given the right high-level guidance, even without a complete system algorithm.
The \textbf{intent} component in \spec{} is designed to provide such guidance.
First, it describes the high-level goal of a function in natural language, e.g., ``successful traversal and insertion'' in \autoref{fig:hoare-spec} 
indicates 
that the target directory is identified through file-tree traversal. 
Second, it allows developers to inject domain-specific knowledge to steer the LLM toward better implementation choices, complementing the correctness guarantees provided by the Hoare logic and invariants.
E.g., when reading a large file extent, a developer can use the intent to suggest a single, bulk I/O operation.
Without this hint, an LLM might generate a correct but inefficient implementation that reads each disk block individually.

\subsection{Modularity Specification}
\label{subs:design-module}


\REVISE{\spec{} addresses the composition challenge, avoiding interface-level dependency errors through a methodology that combines interface contracts with LLM context management.}

\myparagraph{Context-bounded modular synthesis.} 
One of the core innovations in \spec's modularity specification lies in aligning module design with two intrinsic properties of LLM reasoning.
\REVISE{First, strict size constraints ensure each module fits entirely within the model's context window,
enabling holistic analysis during generation.
The specific constraints on module sizes evolve in tandem with improvements in LLM capabilities and context window capacities. 
Take our case study (\textsection\ref{subs:design-put-it-together}) as an example, we limited module sizes to $\leq$500 LoC, which keeps the token consumption for inference generally within approximately 30K tokens.
}
Second, explicit interface contracts govern inter-module dependencies through \emph{Rely-Guarantee} conditions, a formal mechanism adapted from concurrent program verification~\cite{zou2019atomfs}.
This methodology embodies three critical principles:
(1) module implementations must respect their declared dependencies (Rely),
(2) provide guarantees about their behavior (Guarantee),
and (3) compose through logical implication of these contracts.



\input{code-ins-rg}

A critical adaptation occurs in re-imagining rely-guarantee reasoning, originally developed for concurrent thread verification, for modular system synthesis.
Where traditional \emph{rely} conditions specify permissible environment interference for threads,
\spec's \emph{Rely} clauses enumerate a module's assumptions about other components.
Similarly, \emph{Guarantee} clauses replace thread behavior specifications with module interface contracts.
Each module's Rely conditions must be entailed by the Guarantees of its dependencies, enabling compositional correctness through localized synthesis.

In \autoref{fig:code:rely-guarantee}, the module's Rely clause imports critical elements from its dependent modules, e.g.,
the definition of structures, lock/unlock primitives, path traversal function. 
The generated code then exports its own Guarantee, allowing dependents to build upon its functionality without needing to understand internal implementation details.

\myparagraph{Incorporation with external code.}
\REVISE{
\spec{} also supports incorporating external code (e.g., libraries) via the Rely-Guarantee framework.
External code can be integrated by first exposing their \emph{Guarantees}. 
Developers then specify dependencies on these exposed guarantees within the \emph{Rely} clause of their specifications. 
During code generation, the external code is treated as a satisfied dependency, enabling the generated module to correctly invoke external functions without reimplementing them.
}

\input{code-ins-concurrency}

\subsection{Concurrency Specification}
\label{subs:design-concurrency}

Concurrency poses one of the most significant challenges in FS implementation.
A naive approach within the \spec would be to describe concurrent behavior using the existing functionality specification, defining lock states as pre- and post-conditions and outlining the logic in the algorithmic description.
While plausible in theory, our experience reveals that LLMs struggle to correctly synthesize complex concurrent logic from such unified specifications alone.
E.g., when tasked with implementing the notoriously complex \codeword{rename}, state-of-the-art LLMs consistently failed to generate a correct implementation that satisfied the specification.

Our key insight is to decouple concurrent logic from the functional logic.
We separate the concurrency design (e.g., locking) into a standalone \textbf{concurrency specification}. 
This specification is a specialized version of the functionality specification, focusing solely on concurrent behavior.

During code generation, our toolchain first directs the LLM to generate a correct \textit{sequential} version of the code,
focusing only on the primary functionality.
Once this sequential implementation is validated, the toolchain performs a second pass, using the dedicated concurrency specification to instrument the code with the required locking and other concurrent behaviors.
This separation of concerns makes the synthesis task tractable for current LLMs,
allowing them to correctly handle two distinct, complex problems one at a time.

\autoref{fig:code:ins-concurrency} shows an example of the concurrency specification.
The implementation of \texttt{atomfs\_ins} relies on \texttt{locate} and \texttt{check\_ins}. 
In its concurrency specification, the \textbf{Rely} clauses 
capture the locking requirements of these internal functions.
Specifically, since \texttt{locate} requires the \texttt{cur} lock as a precondition, 
while \texttt{atomfs\_ins} itself has no such precondition, 
the generated code must first acquire the lock on \texttt{root\_inum} 
before invoking \texttt{locate(root\_inum, path)}.
This implies that when implementing a module with \spec, 
the LLMs considered not only the functional dependencies between modules but also their locking dependencies.

\input{code-ins-imp}

\myparagraph{Putting it together.}
\autoref{fig:code:atomfs-ins} shows an example of generated \texttt{atomfs\_ins}.
The code fulfils the functionalities 
defined by the pre- and post-conditions, 
correctly invokes functions from other modules in accordance with the modularity specification, 
and handles lock acquisitions/releases properly as specified by the concurrency specification.

\subsection{DAG-Structured Specification Patch}

We introduce a DAG (Directed Acyclic Graph) structured specification patch,
a mechanism that simplifies the evolution of spec-based file systems.
This approach provides a \textbf{self-contained} description of a new feature, explicitly organizes the dependency changes, and defines a clear and consistent workflow for applying the evolution.

\myparagraph{The leaf node: a self-contained change.}
The evolution process begins at a \textbf{leaf node} of the DAG,
which has no dependencies on other patch nodes.
The specification in a leaf node represents a localized, self-contained change, typically within a single module.
It introduces new logic and provides new \codeword{guarantee}s without affecting any other part of the existing system.
A leaf node can also define new data structures or functions that subsequent nodes in the patch will \codeword{rely} on.

\myparagraph{Intermediate nodes: building on guarantees.}
An intermediate node represents a step in the evolution that builds upon previously introduced changes.
Its specification relies on the new \codeword{guarantee}s provided by its child nodes to implement more complex logic.
In turn, this node provides its own, higher-level guarantees, forming a clear dependency chain that progressively constructs the new feature.
\REVISE{
In principle, any modification to a guarantee necessitates corresponding specification adjustments for all modules that depend on it, i.e.,
these modules are included in the patch as intermediate nodes to preclude potential compatibility issues.
}

\myparagraph{The root node: the integration point.}
The root node is the culmination of the evolution, acting as the final integration point.
\REVISE{Unlike the ``tree'' structure, a ``DAG'' structured patch may have multiple root nodes.
Root nodes are characterized by the property that their specification provides semantically unchanged \codeword{guarantees}.}
This equivalence is critical, as it allows the entire chain of new functionality, built up through the DAG, to safely and transparently replace an old implementation.
This substitution serves as the \textbf{``commit point''}, where the evolution is atomically applied to the base system.

\myparagraph{\REVISE{Evolution involving existing modules.}}
\REVISE{
Nodes within a DAG-structured specification patch includes both new modules and modifications to existing ones. 
A modified existing module is treated as a ``new module,'' which can largely reuse the existing specification. 
If a shared component (e.g., \texttt{inode}) is modified, all dependent modules must be regenerated to ``rely'' on the updated version, while other parts of their specifications may not require any modifications. 
These modified modules then replace the original ones after the patch is merged.
}

\myparagraph{The evolution process.}
The unidirectional dependencies between nodes naturally form a DAG,
which dictates the evolution workflow. The process begins with the \spec{} toolchain (\textsection\ref{subs:design-llm-enhancement}) generating code for the leaf nodes.
It then traverses the graph upwards,
synthesizing the implementation for each parent node by leveraging the freshly generated guarantees from its children.
This continues until the toolchain reaches the root node, at which point the new feature is fully implemented and integrated into the file system.

\subsection{The \spec{} Toolchain}
\label{subs:design-llm-enhancement}

We have designed and implemented three LLM-based agents, \compiler{}, \validator{}, and \ass,
\REVISE{
that form the core toolchain for the \spec{} to serve the entire lifecycle of generative file system development, 
The toolchain further effectively mitigates hallucinations in LLMs by addressing two key aspects: 
(1) it helps that LLMs produce outputs aligned with those specifications through the coordination of \compiler{} and \validator{}, and
(2) it helps developers in producing correct specifications by leveraging a strictly structured specification format and the \ass. 
}

\myparagraph{The \compiler{} agent.}
The \compiler{} agent is responsible for translating the high-level, specification-based file system into a low-level C-language implementation (\texttt{ImpFS}) that can be compiled and deployed.
Analogous to a traditional compiler, it processes the source specification and outputs machine-usable code.
Thanks to our modular design, the \compiler{} can operate on one module at a time, confident that the strict enforcement of rely-guarantee conditions will ensure seamless integration into a monolithic whole.

For each module, the \compiler{} employs two primary techniques.
The first is \textbf{two-phase prompting}, which leverages our separation of concerns in the specification.
The agent's first phase generates a correct sequential implementation of the module, focusing only on its core functionality.
In the second phase, it uses the dedicated concurrency specification to instrument this sequential code with the necessary locking and concurrent behaviors.

The second technique is an iterative \textbf{retry-with-feedback} loop used within each phase.
This loop involves two distinct LLM roles: a \texttt{CodeGen} agent generates the implementation, and a separate,
reasoning-focused \texttt{SpecEval} agent reviews the output against the specification.
If the \texttt{SpecEval} agent identifies a flaw, it does not simply report failure; instead,
it generates specific, actionable feedback (e.g., ``The case where function \codeword{foo()} fails is not handled'').
This feedback is then appended to the original prompt, and the \texttt{CodeGen} agent retries.
This refinement cycle continues until the generated code satisfies the specification or an attempt-limit is reached.

This dual-agent design is critical for overcoming LLM hallucination.
In our experience, the code-generation LLM will occasionally produce incorrect implementations,
even with a precise specification.
However, the \texttt{SpecEval} agent effectively detects these flaws.
This is because verifying a solution against a set of rules is a simpler cognitive task than generating the solution from scratch,
and the probability of two distinct models making complementary errors on the same logic is exceedingly low.

\myparagraph{The \validator{} agent.}
The \validator{} agent performs the final, holistic verification of the complete \texttt{ImpFS}.
It combines specification-based review with traditional testing.
First, it re-uses \texttt{SpecEval} logic from \compiler{} to check each fully-generated module against the combined functionality and concurrency specifications.
Second, it integrates with standard software engineering workflows by running a suite of unit and regression tests against the final C-language file system.
This process emulates a modern CI/CD pipeline: the \texttt{SpecEval} component acts as an automated code reviewer verifying adherence to the design,
while the test suite ensures that no existing functionality has regressed.


\myparagraph{The \ass{} agent.}
\ass{} streamlines specification development through a human-in-the-loop process.
A developer provides a draft specification, which the \ass{} first validates and reformats to meet \spec{}'s syntax.
The agent then enters an automated refinement loop.
It repeatedly invokes the \compiler{}; if the \compiler{}'s \texttt{SpecEval} phase identifies a flaw, the \ass{} uses a new \texttt{SpecFine} step to automatically polish the specification based on the feedback before retrying.
This loop concludes in two ways.
On success, the \ass{} provides the developer with the refined specification and the C implementation for validation.
On failure, it returns the last attempted specification annotated with detailed diagnostics, serving as a debug log that guides the developer in resolving the issue.

%% file: code-ins-spec.tex
\begin{figure}
	 \setlength{\belowcaptionskip}{-10pt}
	 \setlength{\abovecaptionskip}{0pt}
	\centering
	\setlength{\columnsep}{0.2cm}
	\begin{minipage}[t]{0.950\linewidth}
		\begin{minted}[
    numbersep=\parindent,
    escapeinside=||,
    linenos,         % Add line numbers
    breaklines,      % Automatically break lines
    fontsize=\fontsize{8pt}{9pt}\selectfont, % Set font size to scriptsize
    tabsize=0        % Set tabsize to 0
    ]{text}
    /* Hoare-style Specification */
    |\textbf{Pre-condition}|:
      path: a NULL-terminated string array
      name: a valid string
    |\textbf{Post-condition}|: 
    |\textbf{Case 1}| Successful traversal and insertion
      - New inode created
      - Entry inserted into target directory  
      - Return 0
    |\textbf{Case 2}| Traversal or insertion failure
      Return -1
    |\textbf{Invariant}|: root_inum always exists  
	\end{minted}
	\end{minipage}
	\caption{Simplified functionality specificaiton for \texttt{atomfs\_ins}}
	\label{fig:hoare-spec}
\end{figure}

%% file: code-ins-rg.tex
\begin{figure}
	 \setlength{\belowcaptionskip}{-10pt}
	 \setlength{\abovecaptionskip}{0pt}
	\centering
	\setlength{\columnsep}{0.2cm}
	\begin{minipage}[t]{0.950\linewidth}
		\begin{minted}[
    numbersep=\parindent,
    escapeinside=||,
    linenos,         % Add line numbers
    breaklines,      % Automatically break lines
    fontsize=\fontsize{8pt}{9pt}\selectfont, % Set font size to scriptsize
    tabsize=0        % Set tabsize to 0
    ]{C}
    |\textbf{[RELY]}|
    |\textbf{Predefined Structures/Functions}|:
    struct inode { ... };
    struct inode* root_inum;
    void lock(struct inode*)
    void unlock(struct inode*)
    struct inode* locate(struct inode* cur, char* path[]) // Traverse path under cur
    void insert(struct inode*, struct inode*, char*) // ...
    int check_ins(struct inode*, char*) // ...

    |\textbf{[GUARANTEE]}|
    |\textbf{Exported Interface}|:
    int atomfs_ins(char*[], char*, int, 
                   unsigned, unsigned)
	\end{minted}
	\end{minipage}
	\caption{\textbf{Rely-Guarantee specifications for \texttt{atomfs\_ins}.} Both rely and guarantee conditions are simplified for clarity.}
	\label{fig:code:rely-guarantee}
\end{figure}

%% file: code-ins-concurrency.tex
\begin{figure}
	\setlength{\belowcaptionskip}{-10pt}
	\setlength{\abovecaptionskip}{0pt}
	\centering
	\setlength{\columnsep}{0.2cm}
	\begin{minipage}[t]{0.950\linewidth}
		\begin{minted}[
    numbersep=\parindent,
    escapeinside=||,
    linenos,         % Add line numbers
    breaklines,      % Automatically break lines
    fontsize=\fontsize{8pt}{9pt}\selectfont, % Set font size to scriptsize
    tabsize=0        % Set tabsize to 0
    ]{text}
    |\textbf{[Rely]}|
    |\textbf{[Locking Specifications of locate]}|
    |\textbf{Pre-condition}|: cur is locked.
    |\textbf{Post-condition}|: suppose return target.
    - if target is NULL, no lock owned.
    - if target is not NULL, only target is owned.
    |\textbf{[Locking Specifications of check\_ins]}|
    |\textbf{Pre-condition:}| cur is locked.
    |\textbf{Post-condition:}|
    - if check_ins returns 0, cur is locked. 
    - if check_ins returns 1, no lock is owned.

    |\textbf{[Locking Specifications of atomfs\_ins]}| 
    |\textbf{Pre-condition:}| no lock is owned. 
    |\textbf{Post-condition:}| no lock is owned.
	\end{minted}
	\end{minipage}
	\caption{\textbf{Concurrency specifications for \texttt{atomfs\_ins}.}}
	\label{fig:code:ins-concurrency}
\end{figure}

%% file: code-ins-imp.tex
\begin{figure}
	 \setlength{\belowcaptionskip}{-10pt}
	 \setlength{\abovecaptionskip}{0pt}
	\centering
	\setlength{\columnsep}{0.2cm}
	\begin{minipage}[t]{0.950\linewidth}
		\begin{minted}[
    numbersep=\parindent,
    escapeinside=||,
    linenos,         % Add line numbers
    breaklines,      % Automatically break lines
    fontsize=\fontsize{8pt}{9pt}\selectfont, % Set font size to scriptsize
    tabsize=0        % Set tabsize to 0
    ]{C}
 int atomfs_ins(char* path[], char* name, ...) {
     lock(root_inum); 
     struct inode* target = locate(root_inum, path);
     if (!target) return -1;
     if (check_ins(target, name) != 0) return -1;
     struct inode* new_inode = malloc_inode(...);
     insert(target, new_inode, name);
     unlock(target); 
     return 0;
 }
	\end{minted}
	\end{minipage}
	\caption{\textbf{LLM-generated \texttt{atomfs\_ins} with \spec.}}
	\label{fig:code:atomfs-ins}
\end{figure}

%% file: design-fs.tex
\input{fs-evolv-case-study}
\section{\sys: A Case for Generative FS}
\label{subs:design-put-it-together}




\subsection{Prototype Implementation}

\myparagraph{Overview.}
To demonstrate the benefits of \spec, we introduce \sys,
a concurrent in-memory file system that runs in userspace via FUSE.
The architecture is based on a prior formally verified file system,
AtomFS~\cite{zou2019atomfs}.
However, rather than porting its C implementation, we undertake a \emph{complete reimplementation} guided by our specification.
Specifically, we implement AtomFS's high-level design with \spec's specifications,
including pre- and post-conditions (we re-use some conditions from AtomFS's formal specifications in Coq),
invariants,
rely/guarantee contracts,
system algorithm descriptions,
and concurrency specification.


\sys is organized into 45 distinct modules distributed across several logical layers,
including file operations, inode management, path traversal, and the POSIX interface.
\sys supports a wide range of standard POSIX calls, such as \codeword{open}, \codeword{read}, and \codeword{rename}.
To validate its functional correctness, we use the \texttt{xfstests}~\cite{xfstest} suite within our \validator{}.
\sys demonstrates a level of correctness equivalent to that of the original AtomFS, failing only 64 out of 754 test cases, all attributable to unimplemented functionality.


\myparagraph{System complexity \REVISE{and comprehensibility.}}
While \sys{} serves as our prototype, it is a non-trivial file system by any measure.
The generated C implementation comprises approximately 4,300 lines of code.
To put this figure into context, we compared it against the 82 file systems in the Linux 6.1.10 kernel.
\sys{} ranks 42nd by line count, surpassing established systems like \texttt{squashfs} and nearly matching the size of \texttt{9pfs}.
\REVISE{We choose C as the programming language as it remains the de facto standard and is most widely used for low-level file system developments.
While high-level languages like Rust may be applicable in the future, 
they present trade-offs between the benefits from their inherent features (e.g., memory safety) and the increased specification complexity (e.g., due to Rust's ownership model).}
\REVISE{Moreover, we observe that the generated code of \sys is highly comprehensible. 
LLMs naturally adhere to standard coding conventions and generate explaining comments, mirroring the style of human engineers. 
}

\myparagraph{Workflow.}
\sys{} features a unique runtime translation workflow, as shown in \autoref{fig:design-overview}-b.
It begins with our LLM-based agents generating C code for each module based on its specification.
A background daemon then compiles the generated code and validates it using the \validator.
Once validated, the compiled components are deployed into the host OS.
To mitigate the latency associated with LLM-based generation, successfully validated module implementations are cached for immediate reuse.
When the specification is updated, regeneration is triggered asynchronously,
allowing the file system to remain fully operational while the new version is prepared in the background.

\myparagraph{\REVISE{Experience for developing \sys{}.}}
\REVISE{
We gain several experiences during the development of \sys{}. 
First, while strict \spec{} adherence effectively ensures accuracy of code generation (\textsection\ref{s:eval-accuracy}), 
simplified specifications (e.g., simplify Rely/Guarantee) can occasionally still result in logically correct code. 
If the objective is to maximize development efficiency, one might consider making a trade-off between generation accuracy and the use of simplified specifications.
Second, to debug \sys{}, we initially locate bugs in the generated C code, 
consistent with debugging methods for human-written code.
Once a bug is localized, the debugging process additionally involves validating the high-level design against the specification. 
In our experience, most issues in \sys{} are diagnosed at the specification level.
}

\subsection{Case Study of Evolutions}
\label{sec:case-study}
One benefit of a specification-based generative FS is the enhancement of the evolvability.
In this section,
we evolve the design of \sys with 10 well-known Ext4's features (\autoref{tab:ext4_features})
using spec patches\footnote{The Appendix supplements the DAG structures of the patches used for specifying these features.}, as a case study.
These ten features are classified into four categories that current \spec can support:
(I) \emph{File structure modification}, which modifies the underlying data structures within the file system;
(II) \emph{Design update for existing operations}, which modifies the specific behavior of existing operations, without altering the intended outcome of these operations;
(III) \emph{New functionalities implementation} with new operations;
(IV) \emph{Hyperparameters or metadata modification}, e.g., changing the FS's block size from 32-bit to 48-bit.
\autoref{tab:ext4_features} presents the specific type of each feature and a brief description of its implementation logic.

\begin{figure}[t]
  \setlength{\belowcaptionskip}{-5pt}
  \setlength{\abovecaptionskip}{-5pt}
  \centering
  \begin{minipage}[t]{0.92\linewidth}
      \centering
      \includegraphics[width=\textwidth]{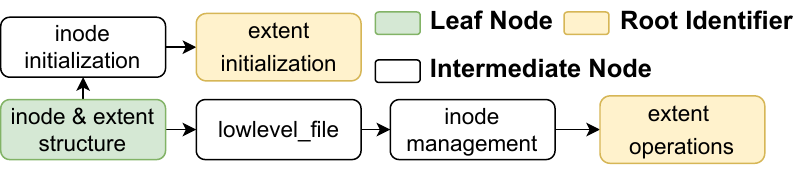}
      \footnotesize
    \end{minipage}
  \caption{\REVISE{\textbf{DAG-Structured patch for implementing ``Extent''.} \textit{For simplicity, a node in the figure may encompass multiple modules of specifications. 
  ``Root Identifier'' indicates that this node is the root (along with its associated logic).}}
  }
  \label{fig:design-feature-prealloc}
\end{figure}

\myparagraph{Example: Extent.}
This feature transforms the file structure from individual blocks into \texttt{extents}. 
Each extent records a segment of contiguous blocks, facilitating sequential reads and writes. 
The modification process (\autoref{fig:design-feature-prealloc}) begins by defining new data structure for \texttt{extent} and \texttt{inode}.
Then, it updates the low-level file operations in \texttt{lowlevel\_file} (and corresponding initializations) to incorporate new extent-based file I/O operations. 
\REVISE{Subsequently, the \texttt{inode\_management} module is modified to invoke these new extent operations.
Since the new \texttt{inode\_management} provides the same guarantee as the original \texttt{inode\_management}, 
the new \texttt{inode\_management} serves as the root node of the entire patch and directly replaces the original one, 
making the complete set of new operations visible to the entire system.
}

%% file: fs-evolv-case-study.tex
\begin{table*}[t]
\centering
\caption{\textbf{Case study of applying Ext4 features to AtomFS}. \textit{``Propose'' and ``Launch`` indicate the year an feature proposed and the year it is merged. 
``Release`` indicates the Linux version with the feature. "Indirect Block" is an exception in the table, as it originates from ext2/3.}}
\label{tab:ext4_features}
\begin{adjustbox}{width=0.95\textwidth}
\footnotesize
\begin{tabular}{@{}llrrp{6cm}cr@{}}
\toprule
\textbf{Type} & \textbf{Feature} & \textbf{Propose} & \textbf{Launch} & \textbf{Brief Description} & \textbf{Release} \\ 
\midrule

\multicolumn{1}{l|}{\multirow{3}{*}{I}}   & Indirect Block (Ext2/3) & / & / & One-to-one block mapping via multi-level pointers & / \\
\multicolumn{1}{l|}{}                     & Extent & 2006 & 2006 & Contiguous block ranges reducing metadata by 50\% & 2.6.19 \\ 
\multicolumn{1}{l|}{}                     & Inline Data & 2011 & 2013 & Store small files in inode's unused space & 3.8 \\ 
\hline

\multicolumn{1}{l|}{\multirow{3}{*}{II}}   & Multi Block Pre-Allocation & 2006 & 2008 & Benefit large files by allocating blocks in groups & 2.6.25 \\ 
\multicolumn{1}{l|}{}                     & Delayed Allocation & 2006 & 2008 & Deferred block allocation for reducing I/O operations & 2.6.27 \\
\multicolumn{1}{l|}{}                     & rbtree for Pre-Allocation & 2022 & 2023 & rbtree to organize the pre-allocated block pool & 6.4 \\ 
\hline

\multicolumn{1}{l|}{\multirow{3}{*}{III}}   & Metadata Checksums & 2011 & 2012 & Checksummed filesystem metadata structures & 3.5 \\ 
\multicolumn{1}{l|}{}                     & Encryption & 2015 & 2015 & Per-directory encryption with low overhead & 4.1 \\ 
\multicolumn{1}{l|}{}                     & Logging(jbd2) & 2006 & 2006 & Journaling support for 64-bit filesystems & 2.6.19 \\
\hline

\multicolumn{1}{l|}{\multirow{1}{*}{IV}}   & Timestamps & 2006 & 2006 & Nanosecond resolution timestamps in inode structure & 2.6.19 \\
\bottomrule
\end{tabular}
\end{adjustbox}
\end{table*}

%% file: eval.tex
\section{Evaluation}
\label{s:eval}

\subsection{Evaluations on Accuracy}
\label{s:eval-accuracy}
\myparagraph{Methodology.}
We first study whether \sys can accurately generate code to achieve self-evolution, 
that is, whether the code generation can accurately conform to the logic described by the specification and accurately meet the functionality of the corresponding file system module.
Our test cases comprise all the modules to implement the complete logic of AtomFS~\cite{zou2019atomfs}. 
Specifically, to evaluate the accuracy, we first define 45 distinct modules within AtomFS and manually author their ground-truth implementations. 
\REVISE{A generated module is considered \textbf{correct} if it (i) passes all functional tests for AtomFS and (ii) is deemed logically equivalent to the ground-truth through manual inspection.}

We implement two versions of baselines based on a few-shot learning approach.
The normal version incorporates a description of the file correspondence logic and the APIs of the dependency modules;
The oracle version not only includes the dependency module's APIs but also integrates the ground-truth code of these modules as part of the prompt.
We test generation accuracy using four LLMs of decreasing capability: Gemini-2.5-Pro~\cite{gemini}, DeepSeek-V3.1 Reasoning~\cite{deepseekai2025deepseekv3technicalreport}, GPT-5-minimal~\cite{gpt5}, and Qwen3-32B~\cite{yang2025qwen3technicalreport}, whose performance is ranked according to the LiveCodeBench leaderboard~\cite{livecodebench-leaderboard}.

\begin{figure}[t]
 \setlength{\belowcaptionskip}{-10pt}
 \setlength{\abovecaptionskip}{0pt}
  \begin{minipage}[t]{0.48\linewidth}
  \centering
  \includegraphics[width=0.98\textwidth]{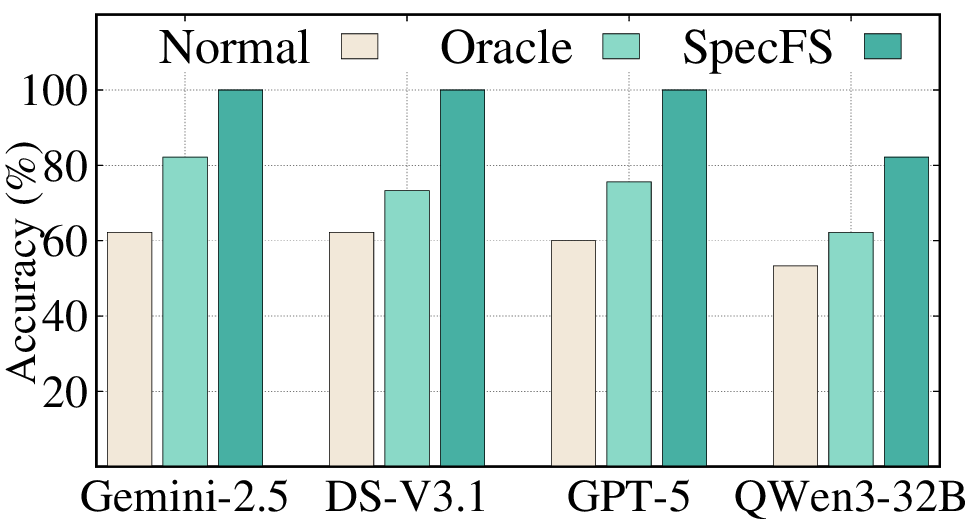}
  \textbf{(a) AtomFS.}
  \end{minipage}
  \begin{minipage}[t]{0.48\linewidth}
    \centering
    \includegraphics[width=0.98\textwidth]{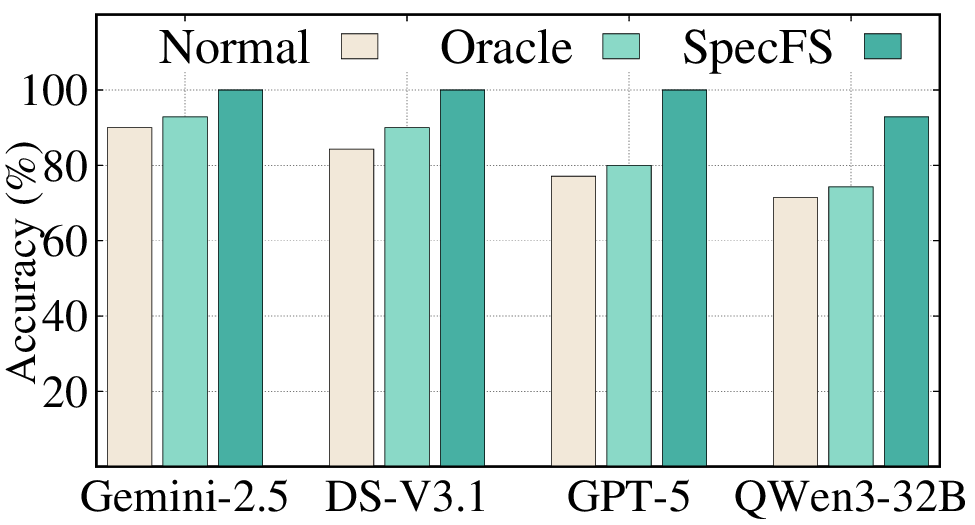}
    \textbf{(b) Features.}
    \end{minipage}
  \caption{\textbf{Accuracy results for implementing AtomFS or new features.} \textit{``GPT-5'' in the figure denotes GPT-5-minimal.}
  }
  \label{fig:eval-accuracy-atomfs}
\end{figure}

\myparagraph{Accuracy results.}
To evaluate accuracy, \sys uses the same specification to generate code for the 45 modules with the four models, 
and the accuracy results are presented in \autoref{fig:eval-accuracy-atomfs}-a. 
We observe that on more powerful models such as Gemini-2.5-Pro and DS-V3.1, 
\sys achieves 100\% accuracy, completely generating all functional modules of \sys. 
In contrast, even when the oracle baseline implementation possesses all contextual code, 
the accuracy of code generation using the most capable Gemini-2.5-Pro is 81.8\%. 
This demonstrates that \sys's specification design significantly enhances code generation accuracy for the file system.

\subsection{Evaluations on Generalizability}

We further evaluate how \spec can support the generation of a wider range of file system logic.
To this end, we evaluate whether \sys can accurately generate the ten features detailed in \autoref{tab:ext4_features}.
These ten new features encompass a total of 64 functional modules, 
and we report their overall accuracy. 
As depicted in \autoref{fig:eval-accuracy-atomfs}-b, \sys consistently exhibits higher generation accuracy across all evaluated models.
This further substantiates the effectiveness of \spec.
Noted that, due to many features being implemented primarily through modifications to existing specifications, and involving less complex concurrency logic, 
the accuracy of implementing features is higher than implementing atomfs from scratch, further reflecting the feasibility of using LLMs for FS evolving.

\myparagraph{Generalizability for various locking methods.}
We evaluate whether the concurrency specifications of \spec are general for various locking methods.
To this end, we utilize the \texttt{dentry\_lookup} operation in the VFS layer of Linux, 
as it exhibits locking requirements at multiple granularities: 
a lock must be acquired for the entire hash list during traversal, 
while individual locks are also required for each dentry upon access. 
In our concurrency specification,
we explicitly designate the use of two distinct locking mechanisms—lock-free RCU for the hash list and spinlocks for individual dentries. 
Our experiments verify that the generated code correctly adheres to the specified concurrency semantics,
successfully producing code that implements multi-granularity and multi-method locking logic as intended.
\REVISE{See details about the specification and the generated code of the two phases in the Appendix.}

\input{tab-ablation}

\subsection{Ablation Study}
We evaluate how \sys's designs effectively improve the accuracy of code generation.
We first divide AtomFS's 45 modules into 40 concurrency-agnostic modules and 5 thread-safe modules, 
and then assess the accuracy for implementing these modules under different design configurations, as shown in \autoref{tab:eval-accuracy-ablation}.
According to the evaluation, 
we observe that although the functionality specification alone is insufficient for correctly generating complex file system (primarily due to interface mismatch), 
when combined with the modularity specification, 
it effectively supports the generation of concurrency-agnostic modules. 
Nevertheless, that is not enough to accurately produce thread-safe modules. 
\sys further addresses this limitation by incorporating a concurrency specification and agent-supported self-validation to achieve correctness and robustness for thread-safe modules.

\input{tab-productivity}

\subsection{Evaluations on Productivity}
We further evaluate how \sys improves the productivity.
We compare the development costs of manual implementation and specification-driven generation for the following two patches:
(1) Supporting the ``extent'' feature for original AtomFS, which requires updating multiple concurrency-agnostic modules; and
(2) Implementing the ``rename'' module of AtomFS, which involves complex locking logic to ensure thread-safety.
We invite two CS master students and two PhD students for the evaluation.

As shown in \autoref{tab:eval-productivity}, \sys improves the programming productivity of the two implementations by 3.0$\times$ and 5.4$\times$.
For implementing patches involving multiple modules, 
\sys's design of DAG-structured specification patch allows for faster identification of all modules requiring modification, 
without the need for source code analysis. 
\sys's concurrency specifications further reduce the complexity of developing sophisticated thread-safe functions.

\begin{figure}[t]
 \setlength{\belowcaptionskip}{-10pt}
 \setlength{\abovecaptionskip}{0pt}
  \begin{minipage}[t]{0.98\linewidth}
    \centering
    \includegraphics[width=0.98\textwidth]{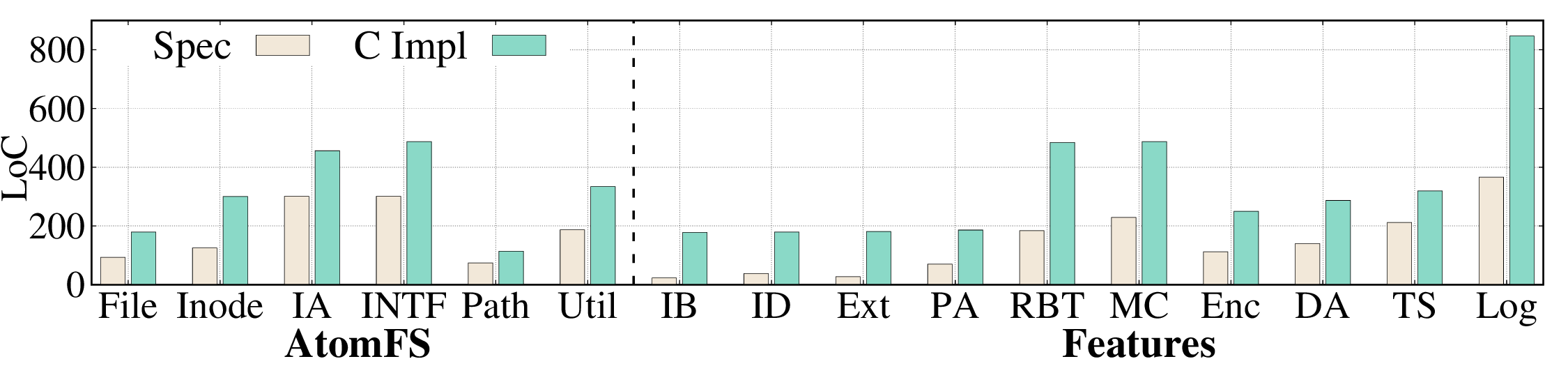}
    \end{minipage}
  \caption{\textbf{Lines of code comparison between specification and source code.} \textit{The results contain six basic logical layers of AtomFS and ten new features in \autoref{tab:ext4_features}.
  The abbreviations are as follows. 
  IA: Interface Auxiliary; INTF: Interface; Util: Utility;
  IB: Indirect Block; ID: Inline Data; Ext: Extent; PA: Pre-Allocation; RBT: rbtree for Pre-Allocation; MC: Metadata Checksums; Enc: Encryption; 
  DA: Delayed Allocation; TS: Timestamps; Log: Logging.}}
  \label{fig:eval-loc}
\end{figure}

\myparagraph{Lines-of-code.}
We conducted a comparative analysis of the lines of code (LoC) between the specifications and the corresponding generated C source code. 
The specifications of AtomFS are categorized by logical layers, each of which may encompass several modules.
The specifications of Features are categorized based on their functional characteristics, each of which corresponds to the features listed in \autoref{tab:ext4_features}. 
The results in \autoref{fig:eval-loc} show that the specification descriptions consistently require fewer lines than their corresponding generated C source code. 
This reduction in LoC implies possibly lower development effort and improved productivity.

\myparagraph{\REVISE{Generation latency.}}
\REVISE{The latency of code generation primarily depends on the inference performance of the LLM itself. 
In our experience, the generation time for \sys{} typically ranges from several minutes to tens of minutes. 
}

\begin{figure}[t]
 \setlength{\belowcaptionskip}{-10pt}
 \setlength{\abovecaptionskip}{5pt}
  \begin{minipage}[t]{0.445\linewidth}
    \centering
    \includegraphics[width=\textwidth]{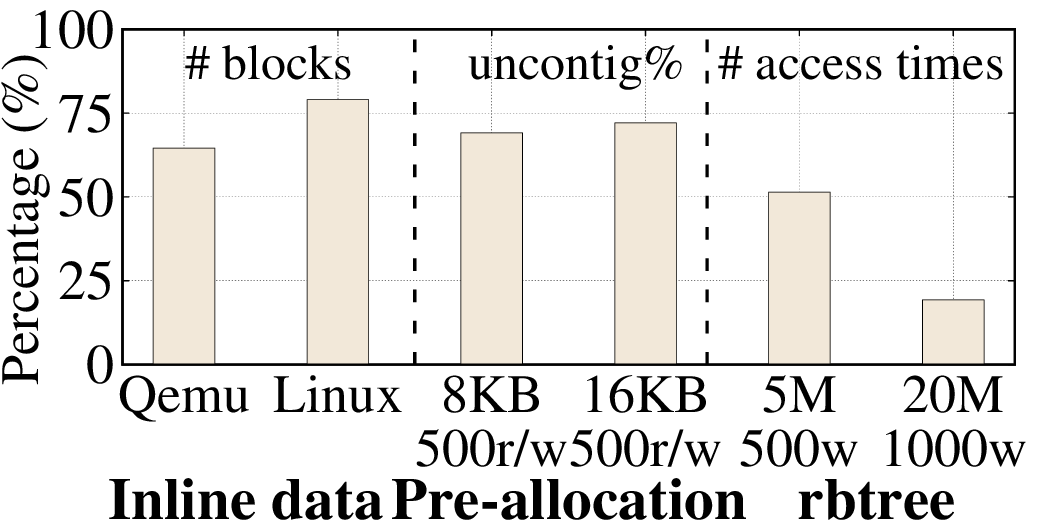}
    \end{minipage}
    \begin{minipage}[t]{0.525\linewidth}
      \centering
      \includegraphics[width=\textwidth]{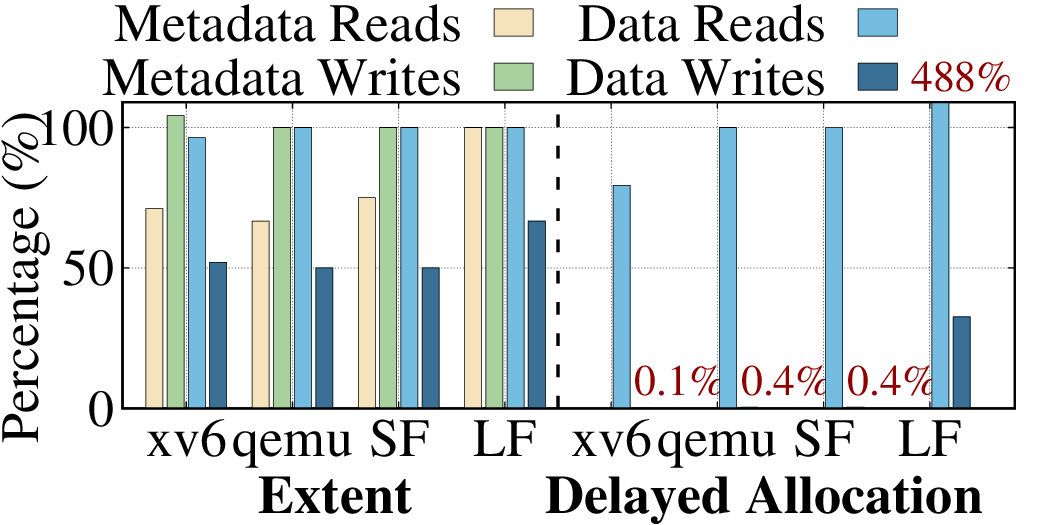}
      \end{minipage}
    \caption{\textbf{Performances improvements with new features implemented by \sys.}
    \textit{The results were normalized before and after optimization, and a lower percentage indicates better performance.
    Test cases in the right are ``xv6 compilation'' (``xv6''), ``copy qemu'' (``qemu''),``small file'' (``SF''), ``large file'' (``LF'').
    The ``SF'' and ``LF'' tests involve various read or write operations on multiple small or large files, respectively representing metadata-intensive and data-intensive workloads.
    The data for "Delayed Allocation" that is either too long or too short is marked in the figure.}}
  \label{fig:eval-perf}
\end{figure}

  

\subsection{Performance Optimizations}
\label{s:eval-perf}
This section evaluates how the new features implemented by \sys (\textsection\ref{sec:case-study}, \autoref{tab:ext4_features}) effectively improve the performance.

\myparagraph{Inline data.}
We first evaluate the compression effect of ``Inline Data'' on file sizes. 
As shown in \autoref{fig:eval-perf}-left, using inline data significantly reduces the file sizes of QEMU~\cite{qemu} and Linux source code~\cite{linux-src}, 
decreasing the required storage capacity by 35.4\% and 21.0\%, respectively.

\myparagraph{Multi-block pre-allocation.}
We evaluate whether the ``Multi-block Pre-allocation'' feature, 
which integrates ``Extent'' and proactively allocates contiguous blocks to form a block pool and subsequently prioritizes drawing blocks from this pool during block allocation, 
increases the sequential read/write ratio of file system operations.
Our microbenchmarks first create a large file and issue random writes to it at fixed page sizes (e.g., 4KB or 8KB). 
Then, we repeatedly apply the following process: a random region within the file is selected, and sequential read or write operations are issued over this range. 
\autoref{fig:eval-perf}-left shows uncontiguous reads and writes ratio drops $\sim$30\% after applying pre-allocation (we regard a read/write operation as sequential if its range falls within a single extent). 
This emphasizes that our pre-allocation optimization enhances the contiguity of data blocks of each file.

\myparagraph{Red-black tree for pre-allocation.}
We evaluate how effectively the transition of the pre-allocation block pool from a linked list to a red-black tree structure reduces block pool access frequency.
Our micro-benchmark first constructs a file with a large block pool by employing a series of write operations exhibiting a specific pattern.
Then, we perform random writes to the file and record the number of accesses to the block pool.
As shown in \autoref{fig:eval-perf}-left, the red-black tree demonstrably reduces block pool access frequency, 
for example, by 80.7\% when performing 1,000 writes on a 20MB file. 
The result also indicates that the benefits of the red-black tree are more pronounced with larger files.

\myparagraph{Extent.}
\autoref{fig:eval-perf}-right shows the proportion of I/O operations (encompassing both metadata and data reads and writes) after applying Extent, relative to those before applying Extent.
The result shows that applying Extent effectively reduces the number of I/O operations, thereby improving the performance of the file system.
It is because an extent data structure represents a sequence of contiguous data blocks.
The read and write operations on this sequence of data blocks are completed in a single I/O operation, 
rather than through multiple individual block-by-block reads and writes.

\myparagraph{Delayed allocation.}
We evaluate how the number of data read and write operations is reduced by ``Delayed Allocation'', 
which prioritizes read and write within a global buffer, and flushes the buffer to the disk in a batch when it reaches the size limits. 
According to the results shown in \autoref{fig:eval-perf}-right, the number of write operations is significantly reduced, with some cases demonstrating an elimination of up to 99.9\% of write operations.
Read operations are also reduced in most cases, while in the cases such as the large file test, the number of data reads increases. 
The reason is that, after applying the feature, data writes no longer directly target the disk; 
instead, data is read into a buffer and write operations are performed within that buffer.
This may introduce additional reads, especially for regular sequential cyclic writes.


%% file: tab-ablation.tex
\begin{table}[t]
  \setlength{\belowcaptionskip}{-5pt}
  \setlength{\abovecaptionskip}{-5pt}
    \caption{\textbf{Ablation study.} \textit{The evaluation uses DeepSeek-V3.1 Reasoning. ``Func'', ``Mod'', ``Con'' means Functionality, Modularity and Concurrency Specifications respectively.}}
    \vspace{-1mm}
    \begin{center}
      \footnotesize
    \renewcommand\arraystretch{1.2}
    \begin{tabular}{c|cccc}
    
    \toprule
    
      \textbf{Modules} & \textbf{Func} & \textbf{+Mod}  & \textbf{+Con} & \textbf{+\validator{}}\\
    \hline
    
    \textbf{\begin{tabular}[c]{@{}c@{}}Concurrency-\\agnostic\end{tabular}} & \begin{tabular}[c]{@{}c@{}}40.00\%\\ (12/40)\end{tabular} & \begin{tabular}[c]{@{}c@{}}100\%\\ (40/40)\end{tabular} & \begin{tabular}[c]{@{}c@{}}100\%\\ (40/40)\end{tabular} & \begin{tabular}[c]{@{}c@{}}100\%\\ (40/40)\end{tabular} \\ 
    \hline
    \textbf{\begin{tabular}[c]{@{}c@{}}Thread-\\safe\end{tabular}}          & \begin{tabular}[c]{@{}c@{}}0\% \\ (0/5)\end{tabular}      & \begin{tabular}[c]{@{}c@{}}0\% \\ (0/5)\end{tabular}    & \begin{tabular}[c]{@{}c@{}}80\%\\ (4/5)\end{tabular}    & \begin{tabular}[c]{@{}c@{}}100\%\\ (5/5)\end{tabular}   \\ 
    \bottomrule

    \end{tabular} \\[-10pt]
    \label{tab:eval-accuracy-ablation}
    \end{center}
\end{table}

%% file: tab-productivity.tex
\begin{table}[t]
  \setlength{\belowcaptionskip}{-5pt}
  \setlength{\abovecaptionskip}{-10pt}
    \caption{\textbf{Productivity improvement.}}
    \begin{center}
      \footnotesize
    \begin{tabular}{c|c|c}
    \toprule
    \textbf{Development Costs} & \textbf{Extent} & \textbf{Rename}  \\
    \hline
    \textbf{Manual} & 4.5h (3.0$\times$) & 13h (5.4$\times$)  \\
    \textbf{Ours} & 1.5h & 2.4h \\
    \bottomrule
    \end{tabular} \\[-10pt]
    \label{tab:eval-productivity}
    \end{center}
\end{table}

%% file: future.tex

\subsection{Limitations and Discussion}
\label{s:future}

\myparagraph{Missing FS features.}
Although \sys represents a successful endeavor in realizing a complex file system using \spec, 
\sys is currently implemented as a user-space file system based on FUSE~\cite{fuse}. 
Consequently, it does not operate in kernel mode and lacks a storage stack, such as direct disk access, nor does it consider crash consistency. 
\REVISE{Our evaluation of \sys primarily focuses on validating the generation methodology and ensuring correctness,
which precludes an apple-to-apple comparison with native kernel-space file systems regarding raw performance metrics such as throughput.
}

\REVISE{We plan to apply \sys to the self-evolution of industrial file systems (e.g., EROFS or Ext4~\cite{gao2019erofs,linux-ext4}),
which is challenging due to the increased engineering complexity.
To address this, we consider a fully formally verified ``AtomFS-Ext4'' unnecessary. 
Instead, we propose developing a ``\sys-Ext4'' directly based on documentation, using these mitigation strategies.
First, instead of specifying the whole state of Ext4, developers can start with a minimal baseline (like Ext2) and incrementally add features to it.
Second, we plan to enhance the \ass{} to automatically bootstrap draft specifications from documentation (e.g., kernel wikis) or even Ext4 source code.
Third, we could adapt methodologies of software engineering to guide the development of specification-constructed (rather than programming-language-constructed) file systems,
e.g., applying methods similar to the Law of Demeter~\cite{demeterlaw} to reduce the coupling between modules.
}

\myparagraph{Push-button verification integration.}
Moreover, although \validator{} enhances the correctness of generated code to some extent through software testing and LLM-based validation, 
a greater potential of \sys lies in the fact that each module is equipped with a ready specification. 
This inherently facilitates integration with push-button verification and similar methodologies, holding the potential to achieve a generative and formally-verified paradigm.

%% file: relat.tex
\section{Related Work}
\label{s:relwk}

We present other related work besides the discussion in \textsection\ref{ssec:opportunity}.

\myparagraph{Domain-specific code generation.}
Some efforts in program synthesis (e.g., SyGus~\cite{sygus}) also focus on automated code generation. 
They may leverage Domain-Specific Languages (DSLs) to precisely articulate code logic,
but face the challenges in scaling to diverse types of scenarios. 
E.g., MegaLibm~\cite{briggs2024megalibm} introduces a novel DSL to construct specifications for mathematical library functions. 
DryadSynth~\cite{ding2024dryadsynth} focuses on bit-vector synthesis building upon the SyGus methodology. 

Other research addresses LLM-based code generation for specific domains outside file systems.
QiMeng Xpiler~\cite{dong2025qimeng} uses LLMs to construct a transcompiler that adapts low-level tensor programs for various hardware platforms. 
Autoverus~\cite{yang2025autoverus} employs LLM-based tools for proving the correctness of Rust code. 
OSVBench and SpecGen~\cite{li2025osvbenchbenchmarkingllmsspecification,ma2025specgenautomatedgenerationformal} address generating formal specifications for a given code snippet. 


\myparagraph{Formal verification methods.}
Many studies verify complex systems, such as operating systems~\cite{gerwin2009sel4}, file systems~\cite{zou2019atomfs,zou2024reffs, chen2016fscq}, or cloud systems~\cite{sun2024anvil}. 
These efforts also involve formal specifications to ascertain the correctness of handcrafted implementations,
but cannot be used diretly for generative file systems. 
We can draw upon these existing specifications when authoring \spec for code generation.

\myparagraph{LLM-assisted development.}
A separate stream of research applies LLMs to mitigate the burden of the file system development. 
For example, WASABI~\cite{stoica2024wasabi} utilizes LLMs to detect intricate retry-related problems in large-scale systems.
SysGPT~\cite{park2025principles} employs LLMs to provide developers with context-aware performance suggestions. 
However, they cannot effectively evolve a file system like \sys.

%% file: concl.tex
\section{Conclusion}
\label{s:conclusion}

This paper presents \spec and \sys, a framework and a case for generative file systems.
Different from traditional paradigms,
\spec shifts the developer's focus from writing low-level code to designing high-level specification, 
and \sys shows the potential benefits on evolvability. 




%% file: ack.tex
\section*{Acknowledgments}  
We are grateful to our shepherd Mai Zheng for his detailed suggestions, which significantly improved the paper. We thank the anonymous FAST reviewers for their constructive feedback.
  This work was supported in part by the National Natural Science Foundation of China (No. 62432010, 62302300 and 62472279), 
the Fundamental and Interdisciplinary Disciplines Breakthrough Plan of the Ministry of Education of China (JYB2025XDXM113),
and the Fundamental Research Funds for the Central Universities.
  Corresponding authors: Dong Du (\url{dd\_nirvana@sjtu.edu.cn}), Yubin Xia (\url{xiayubin@sjtu.edu.cn}), and Haibo Chen (\url{haibochen@sjtu.edu.cn}).

%% file: appendix.tex
\appendix
\clearpage

\section{Artifact Appendix}

\subsection*{Abstract}

This artifact contains the source code and scripts required to reproduce the results presented in the paper. 
The artifact includes the SpecFS filesystem generation pipeline using Large Language Models (LLMs) and end-to-end evaluation workflows. 
It supports reproducing the filesystem generation from high-level specifications, 
executing automated accuracy and performance benchmarks, 
and regenerating the plots reported in the paper.

\subsection*{Scope}

The artifact allows for the validation of the following main claims made in the paper:

\begin{itemize}
    \item \textbf{Accuracy:} The SpecFS filesystem generated by the framework accurately implements the given specifications. Validation is confirmed when the generated filesystem passes all functional tests in the pipeline.
    \item \textbf{Productivity:} The specification descriptions consistently require fewer lines of code than their corresponding generated C source code across evaluated modules, demonstrating improved developer productivity.
    \item \textbf{Performance Optimizations:} The artifact validates that specific optimizations produce measurable performance improvements. 
    The evaluation results allow for a qualitative comparison against the trends reported in the paper.
\end{itemize}

\subsection*{Contents}

The artifact is organized as follows:

\begin{itemize}
    \item \texttt{sysspec/}: Contains the high-level filesystem specifications and the logic for generating the filesystem implementation.
    \item \texttt{eval/}: Includes evaluation artifacts for comparing baselines against optimized versions.
    \item \texttt{data/} \& \texttt{tests/}: Datasets and scripts used by evaluation workloads and validation tests.
    \item \texttt{tools/} \& \texttt{plot/}: Utility scripts for the pipeline and scripts for generating the figures.
    \item \texttt{gen.py}: The main script to run the generation pipeline and functional validation.
    \item \texttt{eval.py}: The main script to execute benchmarks and reproduce evaluation results.
\end{itemize}

\subsection*{Hosting}

The artifact is hosted on GitHub (branch \texttt{main}): \url{https://github.com/LLMNativeOS/specfs-ae}.
The repository includes a detailed \texttt{README.md} file explaining the specific claims and expected results.

\subsection*{Requirements}

\begin{itemize}
    \item \textbf{Operating System:} Linux is required (tested on Debian/Ubuntu).
    \item \textbf{Hardware/Software:} The system must support FUSE (Filesystem in Userspace). Required packages include \texttt{fuse}, \texttt{libfuse-dev}, \texttt{gcc}, \texttt{make}, and \texttt{python3}.
    \item \textbf{Python Environment:} The project uses \texttt{uv} for dependency management.
    \item \textbf{API Access:} An API key is required for the LLM backend. The artifact supports Google AI (Gemini, recommended) or DeepSeek.
\end{itemize}

\section{Case Study: \texttt{dentry\_lookup}}

We take \texttt{dentry\_lookup} function as an representative example to illustrate the specification and the code generated from it. 

\subsection{Specification}

\textbf{Phase 1: Initial Implementation.} Provide a complete C file that implements the \texttt{dentry\_lookup} operation. You can use information from [RELY], [GUARANTEE], and [SPECIFICATION] as described below. Please output only the resulting file.

\myparagraph{[RELY]}
\begin{description}
    \item[Predefined Structures/Functions]
\end{description}

\begin{minted}[
    numbersep=\parindent,
    escapeinside=||,
    % linenos,         % Add line numbers
    breaklines,      % Automatically break lines
    fontsize=\small, % Set font size to scriptsize
    tabsize=0        % Set tabsize to 0
    ]{C}
struct qstr {
    unsigned int hash;
    unsigned int len;
    const unsigned char *name;
};
struct dentry {
    struct qstr d_name;
    struct dentry *d_parent;
    struct hlist_node d_hash;
    atomic_t d_count;
    spinlock_t d_lock;
};
struct hlist_head { /* ... */ };
struct hlist_node { /* ... */ };

struct hlist_head* d_hash(struct dentry* parent, unsigned int hash); 
int memcmp(const void *s1, const void *s2, size_t n);
int d_unhashed(struct dentry* dentry);

#define hlist_entry(ptr, type, member) container_of(ptr,type,member) 
\end{minted}

\myparagraph{[GUARANTEE]}
\begin{description}
    \item[API Compliance:] The function must have the exact signature declared below:
\end{description}

\begin{minted}[
    numbersep=\parindent,
    escapeinside=||,
    % linenos,         % Add line numbers
    breaklines,      % Automatically break lines
    fontsize=\small, % Set font size to scriptsize
    tabsize=0        % Set tabsize to 0
    ]{C}
struct dentry * dentry_lookup(struct dentry * parent, struct qstr * name);
\end{minted}

\myparagraph{[SPECIFICATION]}

\myparagraph{Precondition}: 
\texttt{parent} and \texttt{name} are valid pointers.

\myparagraph{Postcondition}: 
The function's behavior depends on whether a matching dentry is found.
\begin{description}
    \item[Case 1 (Success)] If a child dentry of \texttt{parent} is found with a name that matches \texttt{name}, and this dentry is currently active (not unhashed), then:
    \begin{itemize}
        \item The reference count (\texttt{d\_count}) of the found dentry is incremented.
        \item A pointer to the found dentry is returned.
    \end{itemize}
    \item[Case 2 (Failure)] If no active child dentry of \texttt{parent} with a matching name is found, the function returns \texttt{NULL}.
\end{description}

\myparagraph{System Algorithm:}
\begin{enumerate}
    \item Extract the hash, length, and string from the \texttt{name} parameter.
    \item Use the \texttt{d\_hash} utility to find the correct hash bucket (\texttt{hlist\_head}) associated with the \texttt{parent} dentry.
    \item Iterate through each dentry in the hash bucket in a loop.
    \item For each dentry, perform the following checks:
    \begin{enumerate}[label=\alph*.]
        \item First, compare the hash value with \texttt{name->hash}. If they don't match, skip to the next dentry.
        \item Next, check if \texttt{dentry->d\_parent} is the same as the input \texttt{parent}. If not, skip.
        \item Perform a full name comparison: compare the lengths (\texttt{dentry->d\_name.len} and \texttt{name->len}) and then use \texttt{memcmp} to compare the string content. If the names do not match, skip to the next dentry.
        \item If all checks pass, verify that the dentry is not unhashed using \texttt{d\_unhashed()}.
        \item If it is not unhashed, this is a successful match. Break the loop.
    \end{enumerate}
    \item If a match was found, increment its \texttt{d\_count} and return it. Otherwise, return \texttt{NULL}.
\end{enumerate}

\myparagraph{Phase 2: Concurrency Refinement.} Please refine the above \texttt{dentry\_lookup} function to correctly handle locks for concurrency. Please output only the resulting code. You can rely on the following information.

\myparagraph{[RELY]}
\begin{description}
    \item[Predefined Structures/Functions]
\end{description}

\begin{minted}[
    numbersep=\parindent,
    escapeinside=||,
    % linenos,         % Add line numbers
    breaklines,      % Automatically break lines
    fontsize=\small, % Set font size to scriptsize
    tabsize=0        % Set tabsize to 0
    ]{C}
// Enters an RCU read-side critical section
void rcu_read_lock(void);
// Exits an RCU read-side critical section
void rcu_read_unlock(void);
// Safely dereference a pointer in an RCU critical section
struct hlist_node* rcu_dereference(struct hlist_node* p);
// Acquires a spinlock
void spin_lock(spinlock_t *lock);
// Releases a spinlock
void spin_unlock(spinlock_t *lock);
// Increment a counter atomicly
void atomic_inc(atomic_t *v);
\end{minted}

The locking algorithm for \texttt{dentry\_lookup} has three main components:

\myparagraph{Component 1: RCU-Protected Traversal}

\myparagraph{Precondition:} No RCU lock is held.

\myparagraph{Postcondition:} No RCU lock is held.

\myparagraph{System Algorithm:}
\begin{enumerate}
    \item Before starting the search, call \texttt{rcu\_read\_lock()} to enter an RCU critical section.
    \item After the search is complete (whether a dentry is found or not), call \texttt{rcu\_read\_unlock()} to exit the critical section.
    \item Pointers within the hash list must be accessed via \texttt{rcu\_dereference()} to ensure the most up-to-date value is read safely.
\end{enumerate}

\myparagraph{Component 2: Per-Dentry Spinlock}

\myparagraph{Precondition:} The code is inside an RCU read-side critical section.

\myparagraph{Postcondition:} In all code paths after the lock has been acquired (e.g., after a successful match, after a failed re-check, or after a failed name comparison), the \texttt{dentry->d\_lock} is released.

\myparagraph{System Algorithm:}
\begin{enumerate}
    \item Inside the hash list traversal loop, after finding a candidate dentry (i.e., its hash matches), acquire its individual spinlock.
    \item \textbf{Critical Re-check:} Immediately after acquiring the lock, re-check that \texttt{dentry->d\_parent} still points to the \texttt{parent} you are searching under. If the parent no longer matches, release the lock and continue to the next dentry.
    \item While holding the spinlock, perform the full, definitive name comparison. The lock ensures that \texttt{dentry->d\_name} will not change.
    \item If the name matches and the dentry is not unhashed, increment its reference count. This action must be performed before releasing the lock.
\end{enumerate}

\myparagraph{Component 3: Atomic counter Incrementation}

\myparagraph{Precondition:} No lock is held on the counter.

\myparagraph{Postcondition:} The counter of the dentry is atomicly incremented. No lock is held on the counter.

\subsection{Code Snippet}

\myparagraph{\REVISE{Phase 1: Initial Implementation.}}

\begin{minted}[
    numbersep=\parindent,
    escapeinside=||,
    % linenos,         % Add line numbers
    breaklines,      % Automatically break lines
    fontsize=\small, % Set font size to scriptsize
    tabsize=0        % Set tabsize to 0
    ]{C}
struct dentry * dentry_lookup(struct dentry * parent, struct qstr * name) {
    unsigned int hash = name->hash;
    unsigned int len = name->len;
    const unsigned char *str = name->name;

    struct hlist_head *head = d_hash(parent, hash);
    struct hlist_node *node;
    struct dentry *found = NULL;

    for (node = head->first; node != NULL; node = node->next) {
        struct dentry *dentry = hlist_entry(node, struct dentry, d_hash);

        if (dentry->d_name.hash != hash)
            continue;

        if (dentry->d_parent != parent)
            continue;

        if (dentry->d_name.len != len)
            continue;
        
        if (memcmp(dentry->d_name.name, str, len) != 0)
            continue;

        if (d_unhashed(dentry))
            continue;

        found = dentry;
        break;
    }

    if (found) {
        found->d_count.counter++;
        return found;
    }

    return NULL;
}
\end{minted}

\myparagraph{\REVISE{Phase 2: Refined Implementation.}}

\begin{minted}[
    numbersep=\parindent,
    escapeinside=||,
    % linenos,         % Add line numbers
    breaklines,      % Automatically break lines
    fontsize=\small, % Set font size to scriptsize
    tabsize=0        % Set tabsize to 0
    ]{C}
struct dentry * dentry_lookup(struct dentry * parent, struct qstr * name) {
    struct hlist_node *node;
    struct dentry *found = NULL;
    unsigned int hash = name->hash;

    rcu_read_lock();

    struct hlist_head *head = d_hash(parent, hash);

    for (node = rcu_dereference(head->first); node != NULL;
         node = rcu_dereference(node->next)) {
        
        struct dentry *dentry = hlist_entry(node, struct dentry, d_hash);

        if (dentry->d_name.hash != hash)
            continue;

        spin_lock(&dentry->d_lock);

        if (dentry->d_parent != parent) {
            spin_unlock(&dentry->d_lock);
            continue;
        }

        if (dentry->d_name.len != name->len ||
            memcmp(dentry->d_name.name, name->name, name->len) != 0) {
            spin_unlock(&dentry->d_lock);
            continue;
        }

        if (d_unhashed(dentry)) {
            spin_unlock(&dentry->d_lock);
            continue;
        }

        atomic_inc(&dentry->d_count);
        spin_unlock(&dentry->d_lock);
        
        found = dentry;
        break;
    }

    rcu_read_unlock();

    return found;
}
\end{minted}

\section{DAG-Structured Patches for Features}

\begin{figure*}[t!]
    \vspace*{-13cm} 
    \centering
    \begin{minipage}[t]{0.98\linewidth}
        \centering
        \includegraphics[width=\textwidth]{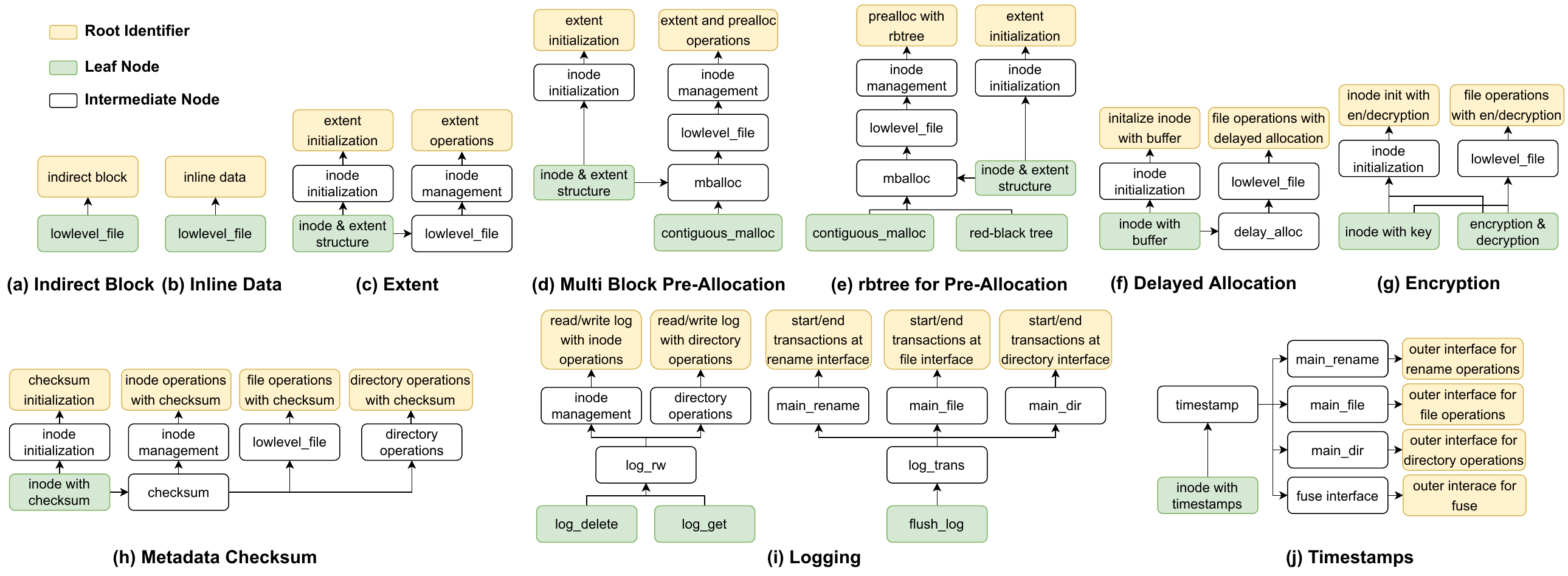}
        \footnotesize
      \end{minipage}
    \caption{\textbf{Features implemented by SpecFS.} \textit{For simplicity, a node in the figure may encompass multiple modules of specifications. 
    ``Root Identifier'' indicates that this node is the root node, along with its associated logic.}
    }
    \label{fig:patches}
\end{figure*}

\autoref{fig:patches} presents the DAG structures of patches used to implement the ten new features of SpecFS.

